\documentclass{jfm}
\usepackage{graphicx, color}
\usepackage{epstopdf, epsfig}
\usepackage{amsmath}

\usepackage{floatrow}
\usepackage{subfig}
\usepackage{multirow}

\usepackage{subfig}

\usepackage{cleveref}
\crefname{section}{section}{sections}
\crefname{subsection}{subsection}{subsections}
\crefname{figure}{figure}{figures}
\crefname{table}{table}{tables}
\crefname{equation}{}{}
\Crefname{section}{Section}{Sections}
\Crefname{subsection}{Subsection}{Subsections}
\Crefname{figure}{Figure}{Figures}
\Crefname{table}{Table}{Tables}

\newcommand{\beq}{\begin{equation}}
\newcommand{\eeq}{\end{equation}}
\newcommand{\lan}{\left\langle}
\newcommand{\ran}{\right\rangle}

\newcommand\Nu{\mbox{\textit{Nu}}}
\newcommand\Ra{\mbox{\textit{Ra}}}
\newcommand{\cO}{\mathcal{O}}

\newcommand\solidrule[1][15pt]{\rule[0.5ex]{#1}{1pt}}
\newcommand\dashedrule{\mbox{%
	\solidrule[3pt]\hspace{3pt}\solidrule[3pt]\hspace{3pt}\solidrule[3pt]}}

\shorttitle{Steady Rayleigh--B\'enard convection between stress-free boundaries}
\shortauthor{B. Wen, D. Goluskin, M. LeDuc, G. P. Chini and C. R. Doering}

\title{Steady Rayleigh--B\'enard convection between stress-free boundaries}

\author{
\mbox{Baole Wen}\aff{1}\corresp{\email{baolew@umich.edu}},
\mbox{David Goluskin}\aff{2},
\mbox{Matthew LeDuc}\aff{3},
\mbox{Gregory P. Chini}\aff{4,5},
 \and
\mbox{Charles R. Doering}\aff{1,3,6}}

\affiliation{
\aff{1}Department of Mathematics, University of Michigan, Ann Arbor MI 48109-1043
\aff{2}Department of Mathematics \& Statistics, University of Victoria, Victoria, BC V8P 5C2
\aff{3}Department of Physics, University of Michigan, Ann Arbor, MI 48109-1040
\aff{4}Program in Integrated Applied Mathematics, University of New Hampshire, Durham NH 03824
\aff{5}Department of Mechanical Engineering, University of New Hampshire, Durham NH 03824
\aff{6}Center for the Study of Complex Systems, University of Michigan, Ann Arbor MI 48109-1042}

\begin{document}

\maketitle

\begin{abstract} 
Steady two-dimensional Rayleigh--B\'enard convection between stress-free isothermal boundaries is studied via numerical computations. We explore properties of steady convective rolls with aspect ratios $\pi/5\le\Gamma\le4\pi$, where $\Gamma$ is the width-to-height ratio for a pair of counter-rotating rolls, over eight orders of magnitude in the Rayleigh number, $10^3\le\Ra\le10^{11}$, and four orders of magnitude in the Prandtl number, $10^{-2}\le\Pran\le10^2$. At large $\Ra$ where steady rolls are dynamically unstable, the computed rolls display $\Ra \rightarrow \infty$ asymptotic scaling. In this regime, the Nusselt number $\Nu$ that measures heat transport scales as $\Ra^{1/3}$ uniformly in $\Pran$. The prefactor of this scaling depends on $\Gamma$ and is largest at $\Gamma \approx 1.9$. The Reynolds number $\Rey$ for large-$\Ra$ rolls scales as $\Pran^{-1} \Ra^{2/3}$ with a prefactor that is largest at $\Gamma \approx 4.5$. All of these large-$\Ra$ features agree quantitatively with the semi-analytical asymptotic solutions constructed by \citet{ChiniCox2009}. Convergence of $\Nu$ and $\Rey$ to their asymptotic scalings occurs more slowly when $\Pran$ is larger and when $\Gamma$ is smaller.
\end{abstract}


\section{Introduction}


Natural convection is the buoyancy-driven flow resulting from unstable density variations, typically due to thermal or compositional inhomogeneities, in the presence of a gravitational field. It remains the focus of experimental, computational, and theoretical research worldwide, in large part because buoyancy-driven flows are central to engineering heat transport, atmosphere and ocean dynamics, climate science, geodynamics, and stellar physics. Rayleigh--B\'enard convection, in which a layer of fluid is confined between isothermal horizontal boundaries with the higher temperature on the underside \citep{Rayleigh1916} is studied extensively as a relatively simple system displaying the essential phenomena. Beyond the importance of buoyancy-driven flow in applications, Rayleigh's model has served for more than a century as a primary paradigm of nonlinear physics \citep{MalkusVeronis1958}, complex dynamics \citep{Lorenz1963}, pattern formation \citep{NewellWhitehead1969}, and turbulence \citep{Kadanoff2001}.

A central feature of Rayleigh--B\'enard convection is the Nusselt number $\Nu$, the factor by which convection enhances heat transport relative to conduction alone. A fundamental challenge for the field is to understand how $\Nu$ depends on the dimensionless control parameters: the Rayleigh number $\Ra$, which is proportional to the imposed temperature difference across the layer, the fluid's Prandtl number $\Pran$, and geometric parameters like the domain's width-to-height aspect ratio $\Gamma$. \citet{Rayleigh1916} studied the bifurcation from the static conduction state (where $\Nu =1$) to convection (where $\Nu >1$) when $\Ra$ exceeds a $\Pran$-independent finite value. In the strongly nonlinear large-$\Ra$ regime relevant to many applications, convective turbulence is characterized by chaotic plumes that emerge from thin thermal boundary layers and stir a statistically well-mixed bulk. Power-law behavior where $\Nu$ scales like $\Pran^\beta \Ra^\gamma$ is often presumed for heat transport in the turbulent regime, but heuristic theories---i.e., physical arguments relying on uncontrolled approximations---yield various predictions for the scaling exponents. Rigorous upper bounds on $\Nu$ derived from the equations of motion place restrictions on possible asymptotic exponents but do not imply unique values. Meanwhile, direct numerical simulations (DNS) and laboratory experiments designed to respect the approximations employed in Rayleigh's model have produced extensive data on $\Nu$ over wide ranges of $\Ra$, $\Pran$, and $\Gamma$.  Even so, consensus regarding the asymptotic large-$\Ra$ behavior of $\Nu$ remains to be achieved \citep{Chilla2012, Doering2020}.

\begin{figure}[t!]
\vspace{0.20in}
\begin{minipage}[t]{0.03\textwidth}
\vspace{0pt}
(a)
\end{minipage}
\begin{minipage}[t]{0.46\textwidth}
\vspace{0pt}
\includegraphics[height=1.16in]{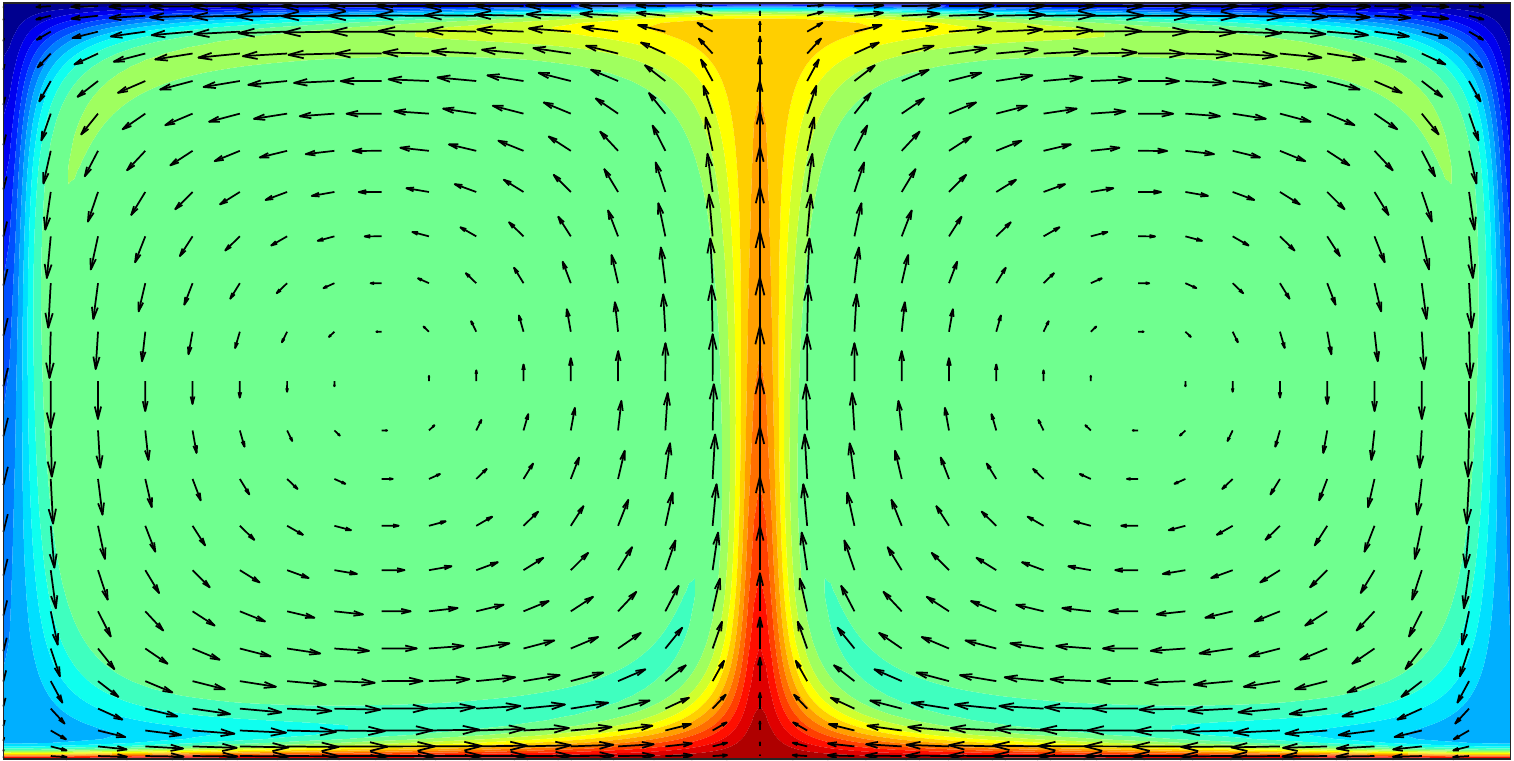}
\end{minipage}\hfill
\begin{minipage}[t]{0.03\textwidth}
\vspace{0pt}
(b)
\end{minipage}
\begin{minipage}[t]{0.46\textwidth}
\vspace{0pt}
\includegraphics[height=1.16in]{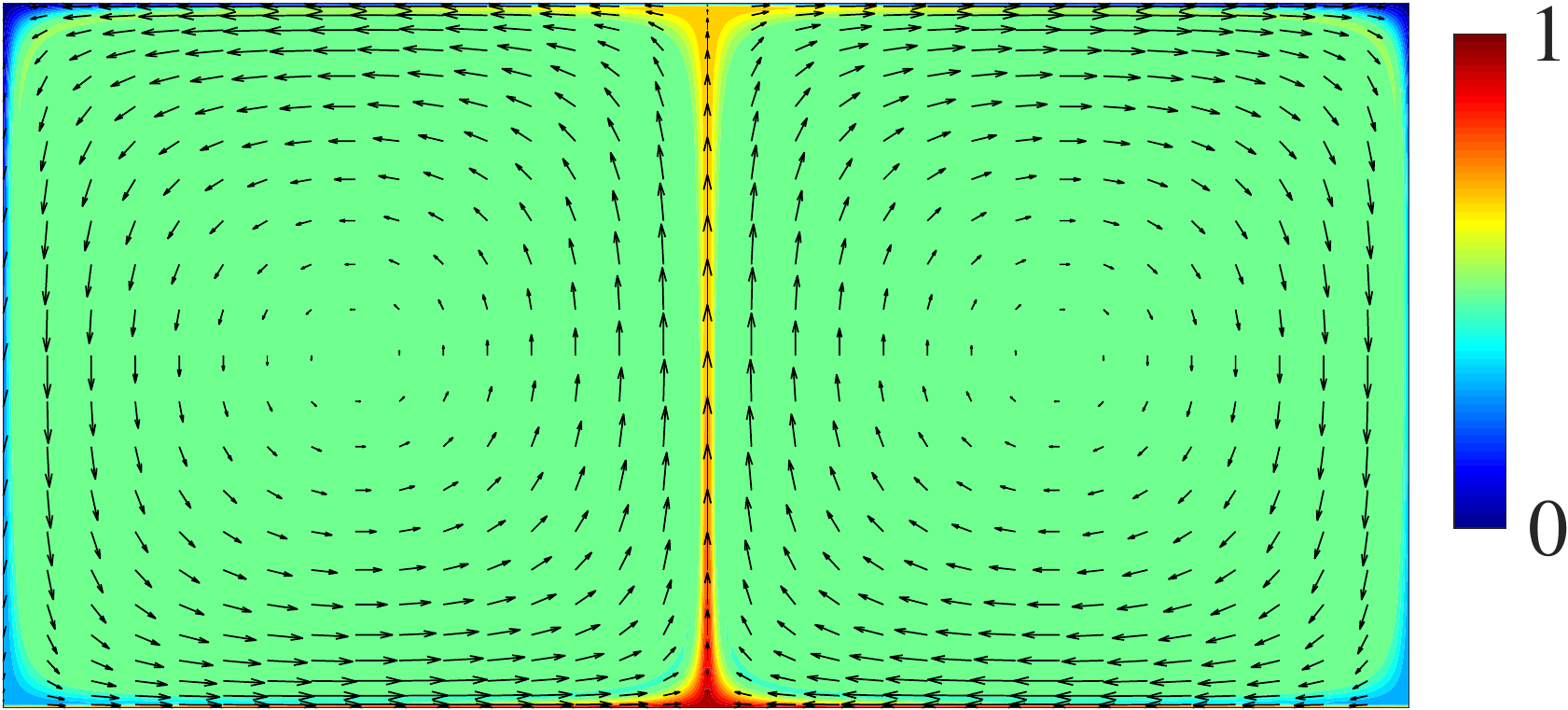}
\end{minipage}
\caption{\label{fig: rolls}Steady convective rolls between stress-free boundaries for (a) $Ra=10^6$ and (b) $Ra=10^{8}$, with $Pr=1$ and a horizontal period that is twice the layer height. Color represents dimensionless temperature and arrows indicate the velocity vector field. As $\Ra\to\infty$, the temperature field develops an isothermal core while the thermal boundary layers and plumes become thinner and the velocity field converges to a $\Ra$-independent pattern that lacks boundary layers.}
\end{figure}

In addition to the turbulent convection generally observed at large $\Ra$, there are much simpler steady solutions to the equations of motion such as the pair of steady counter-rotating rolls shown in \cref{fig: rolls}. Steady coherent flows are not typically seen in large-$\Ra$ simulations or experiments because they are dynamically unstable. Nonetheless, they are part of the global attractor for the infinite-dimensional dynamical system defined by Rayleigh's model, and recent results suggest that steady rolls may be one of the key coherent states comprising the `backbone' of turbulent convection. In the case of no-slip top and bottom boundaries \citet{Waleffe2015} and \citet{Sondak2015} found that, over the range of Rayleigh numbers they explored, two-dimensional (2D) steady rolls display $\Nu$ values very close to those of three-dimensional (3D) convective turbulence, provided that the horizontal period of the rolls is tuned to maximize $\Nu$ at each value of~$\Ra$.

Here we report computations of steady 2D convective rolls in the case of stress-free top and bottom boundaries. We have carried out computations using spectral methods over eight orders of magnitude in $\Ra$, four orders of magnitude in $\Pran$, and more than an order of magnitude in the aspect ratio $\Gamma$, defined as the width-to-height ratio of a pair of rolls. As in the no-slip case, our steady states share many features with time-dependent simulations between stress-free boundaries \citep{Paul2012, Wang2020}. Moreover, the results verify predictions about the $Ra\rightarrow\infty$ limit made by \citet{ChiniCox2009} who extended an approach initiated by \citet{Robinson1967} to construct matched asymptotic approximations of steady rolls between stress-free boundaries. In particular, our computations agree quantitatively with the asymptotic prediction that $\Nu =\cO(\Ra^{1/3})$ uniformly in $\Pran$ with a $\Gamma$-dependent prefactor that assumes its maximum value at $\Gamma \approx 1.9$, and with corresponding asymptotic predictions about the Reynolds number that are derived in the Appendix.
The rest of this paper is organized as follows.
The equations governing Rayleigh--B\'enard convection and our numerical scheme for computing steady solutions are outlined in \S\,\ref{sec: framework}. The computational results are presented in \S\,\ref{sec: results}, followed by further discussion in~\S\,\ref{sec: con}.

\section{\label{sec: framework}Governing equations and computational methods}

The Boussinesq approximation to the Navier--Stokes equations used by \citet{Rayleigh1916} to model convection in a 2D fluid layer are, in dimensionless variables,
\begin{subequations} 
\label{eq: bouss}
\begin{align}
\partial_t\mathbf u + \mathbf u \bcdot \bnabla \mathbf u  &= 
-\bnabla p + \Pran \nabla^2 \mathbf u + \Pran \Ra\,T \mathbf{\hat z}, \label{eq: u} \\
\bnabla \bcdot \mathbf u &= 0, \label{eq: inc} \\
\partial_t T + \mathbf u \bcdot \bnabla T& = \nabla^2 T, \label{eq: T}
\end{align}
\end{subequations}
where $\mathbf{u} = u\mathbf{\hat x} + w\mathbf{\hat z}$ is the velocity, $p$ is the pressure, and $T$ is the temperature. The system has been nondimensionalized using the layer thickness $h$, the thermal diffusion time $h^2/\kappa$, where $\kappa$ is the thermal diffusivity, and the temperature drop $\Delta$ from the bottom boundary to the top one.

The dimensionless spatial domain is $(x,z)\in[0,\Gamma]\times[0,1]$, and all dependent variables are taken to be $\Gamma$-periodic in $x$.
At the lower ($z=0$) and upper ($z=1$) boundaries, the temperature satisfies isothermal conditions while the velocity field satisfies no-penetration and stress-free boundary conditions:
\begin{align}
T|_{z=0}=1 \ \text{and} \  T|_{z=1} = 0, \qquad w|_{z=0,1}=0, \qquad \partial_z u|_{z=0,1}=0.
\label{eq: BCs}
\end{align}
The three dimensionless parameters of the problem are the aspect ratio $\Gamma$, the Prandtl number $\Pran=\nu/\kappa$, where $\nu$ is the kinematic viscosity, and the Rayleigh number $Ra=g\alpha \Delta h^3/\nu\kappa$, where $-g\mathbf{\hat z}$ is the gravitational acceleration vector and $\alpha$ is the thermal expansion coefficient. A single pair of the steady rolls computed here fits in the domain, meaning the aspect ratio of the pair is $\Gamma$ while that of each individual roll is $\Gamma/2$.

The static conduction state, for which $\mathbf{u}=\mathbf0$ and $T=1-z$, solves \cref{eq: bouss,eq: BCs} at all parameter values. \citet{Rayleigh1916} showed that rolls vertically spanning the layer with aspect ratio $\Gamma$ bifurcate supercritically from the conduction state as $\Ra$ increases past
\beq
Ra_c(k) = \frac{(k^2+\pi^2)^3}{k^2},
\eeq
where $k=2\pi/\Gamma$ is the wavenumber of the fundamental period of the domain.
The conduction state is absolutely stable if $\Ra<\Ra_c(k)$ for all $k$ admitted by the domain; see, e.g., \citet{Goluskin2015a}.

The Nusselt number is defined as the ratio of total mean heat flux in the vertical direction to the flux from conduction alone: 
\beq
\Nu = 1+ \langle wT\rangle,
\label{eq: Nu}
\eeq
where $w$ and $T$ are dimensionless solutions of \cref{eq: bouss} and $\langle \cdot \rangle$ indicates an average over space and infinite time. (For steady states the time average is not needed.) The governing equations imply the equivalent expressions
\beq
\Nu = \langle |\bnabla T|^2\rangle = 1+ \frac{1}{Ra} \langle |\bnabla u|^2 + |\bnabla w|^2 \rangle,
\label{eq: NuEquiv}
\eeq
the latter of which self-evidently ensures $\Nu > 1$ for all sustained convection. Another emergent measure of the intensity of convection is the bulk Reynolds number defined using the dimensional root-mean-squared velocity $U_{rms}$, which in terms of dimensional quantities is $\Rey ={U_{rms} h}/{\nu}$.  We choose our reference frame such that $\lan u\ran=0$, so in dimensionless terms
\beq
\Rey = \frac{1}{\Pran}\langle u^2+w^2\rangle^{1/2}.
\label{eq: Re}
\eeq

We compute steady ($\partial_t = 0$) solutions of \cref{eq: bouss} using a vorticity--stream function formulation,
\begin{subequations}
\label{eq: steadybouss}
\begin{align}
\partial_z\psi\partial_x\omega - \partial_x\psi\partial_z\omega  &= 
\Pran \nabla^2 \omega + \Pran \Ra\partial_x \theta, \label{eq: omega} \\
\nabla^2 \psi &= -\omega, \label{eq: psi}\\
\partial_z\psi\partial_x\theta - \partial_x\psi\partial_z\theta & = -\partial_x\psi + \nabla^2 \theta, \label{eq: theta}
\end{align}
\end{subequations}
where the stream function $\psi$ is defined by $\mathbf{u} = \mathbf{\hat x} \partial_z \psi - \mathbf{\hat z} \partial_x \psi$, the (negative) scalar vorticity is $\omega=\partial_x w- \partial_z u = -\nabla^2 \psi$, and $\theta$ is the deviation of the temperature field $T$ from the conduction profile $1-z$. The boundary conditions used in our computations are that $\psi$, $\nabla^2\psi$, and $\theta$ vanish on both boundaries. The latter two conditions follow from the stress-free and fixed-temperature conditions, respectively. Impenetrability of the boundaries implies that $\psi$ is constant on each boundary, and choosing the reference frame where $\lan u \ran=0$ requires these constants to be identical. Their value can be fixed to zero since translating $\psi$ by a constant does not affect the dynamics. 

We solve the time-independent equations \cref{eq: steadybouss} numerically using a Newton--GMRES (generalised minimal residual) iterative scheme.  Starting with an initial iterate $(\omega^0,\psi^0,\theta^0)$ that does not exactly solve \cref{eq: steadybouss}, each iteration of the Newton's method applies a correction until the resulting iterates have converged to a solution of \cref{eq: steadybouss}. Following \citet{Wen2015JFM} and \citet{WenChini2018JFM}, the linear partial differential equations for the corrections are
\begin{subequations}
\label{eq: Omega_Psi_Theta}
\begin{align}
(\Pran\nabla^2 - \psi_z\partial_x + \psi_x\partial_z)^i\triangle^{\omega} +
 (-\omega_x\partial_z + \omega_z\partial_x)^i\triangle^{\psi} +
 \Ra\Pran\partial_x\triangle^{\theta} &= -{F^\omega_{res}}^i, \label{eq: Delta_omega}\\
\triangle^{\omega} + \nabla^2\triangle^{\psi} &= -{F^\psi_{res}}^i, \label{eq: Delta_psi}\\
(-\partial_x + \theta_z\partial_x - \theta_x\partial_z)^i\triangle^{\psi} +
 (\nabla^2 - \psi_z\partial_x + \psi_x\partial_z)^i\triangle^{\theta} &= -{F^\theta_{res}}^i,\label{eq: Delta_theta}
\end{align}
\end{subequations}
where the superscript $i$ denotes the $i^{th}$ Newton iterate, the corrections are defined as
\begin{eqnarray}
    \triangle^{\omega} = \omega^{i+1} - \omega^{i}, \quad \triangle^{\psi} = \psi^{i+1} - \psi^{i}, \quad \triangle^{\theta} = \theta^{i+1} - \theta^{i} \label{Correction}
\end{eqnarray}
and vanish on the boundaries, and
\begin{subequations}
\label{eq: res}
\begin{align}
F^\omega_{res} &= \Pran\nabla^2\omega + \Ra\Pran\theta_x - \psi_z\omega_x + \psi_x\omega_z,\\
F^\psi_{res} &= \nabla^2\psi + \omega,\\
F^\theta_{res} &= \nabla^2 \theta - (\psi_z \theta_x - \psi_x \theta_z + \psi_x)
\end{align}
\end{subequations}
are the residuals of the nonlinear steady equations \cref{eq: steadybouss}. We simplify the implementation by setting $F^\psi_{res}=0$, in which case $\triangle^{\psi}$ can be obtained by solving $\nabla^2\triangle^{\psi} =  -\triangle^{\omega}$ for a given $\triangle^{\omega}$.  After this simplification, the pair \cref{eq: Delta_omega,eq: Delta_theta} can be solved simultaneously for $\triangle^{\omega}$ and $\triangle^{\theta}$.

For each iteration of Newton's method, we solve \cref{eq: Delta_omega} and \cref{eq: Delta_theta} iteratively using the GMRES method \citep{Trefethen1997}. The spatial discretization is spectral, using a Fourier series in $x$ and a Chebyshev collocation method in $z$ \citep{Trefethen2000}.  The $\nabla^2$ operator is used as a preconditioner to accelerate convergence of the GMRES iterations. The roll states of interest have centro-reflection symmetries (cf.\ \cref{fig: rolls}),
\begin{equation}
\label{eq: sym}
[\omega, \psi, \theta](x, z) = [\omega, \psi, -\theta](\Gamma/2 - x, 1- z), \;\;
[\omega, \psi, \theta](x, z) = [-\omega, -\psi, \theta](\Gamma - x, z),
\end{equation}
which allow the full fields to be recovered from their values on one quarter of the domain, so we encode these symmetries to reduce the number of unknowns. The GMRES iterations are stopped once the $L^2$-norm of the relative residual of (\ref{eq: Omega_Psi_Theta}$a,c$) is less than $10^{-2}$, and the Newton iterations are stopped once the $L^2$-norm of the relative residual of (\ref{eq: steadybouss}$a,c$) is less than $10^{-10}$.  For $\Ra$ not far above the critical value $\Ra_c(k)$, convergence to rolls of period $\Gamma=2\pi/k$ is accomplished by choosing the initial iterate with $\omega^0 = -\sqrt{\Ra\Pran}\sin(\pi z)\sin(k x)$ and $\theta^0 = -0.1[\sin(2\pi z) +\sin(\pi z)\cos(k x)]$.  For each $Pr$, results from smaller $\Ra$ (or $\Gamma$) are used as the initial iterate for larger $\Ra$ (or $\Gamma$).

\section{\label{sec: results}Results}

We computed steady rolls over a wide range of $\Ra$ starting just above the value $\Ra_c(k)$ at which the rolls bifurcate from the conduction state and ranging up to $10^9$ or higher depending on the other parameters. Computations were carried out for $\Pran=10^{-2}, 10^{-1},1,10,10^2$ and a range of values of $\Gamma$ such that the fundamental wavenumber $k=2\pi/\Gamma$ lies in $1/2\le k\le10$. Data for all the $\Gamma=2$ cases are tabulated in the Supplementary Material.

Our computations reach sufficiently large $\Ra$ to show clear asymptotic scalings of bulk quantities:
\begin{align}
\Nu\sim c_n(k)\Ra^{1/3} \quad \mbox{and} \quad \Rey\sim c_r(k)\Pran^{-1}\Ra^{2/3} \quad \mbox{as} \quad \Ra \rightarrow \infty.
\label{eq: scalings}
\end{align}
Both of these scalings are predicted by the asymptotic analysis of \citet{ChiniCox2009}, although only the $\Nu$ scaling was stated explicitly there. \citeauthor{ChiniCox2009} gave an asymptotic prediction for the prefactor $c_n(k)$ but not for $c_r(k)$. Using their asymptotic approximations for the stream function and vorticity within each convection roll, we derived an expression for $c_r(k)$ in terms of $c_n(k)$ that is presented in the Appendix.

\begin{figure}
\vspace{0.20in}
\begin{minipage}[t]{0.025\textwidth}
\vspace{0pt}
(a)
\end{minipage}
\begin{minipage}[t]{0.46\textwidth}
\vspace{0pt}
\includegraphics[height=1.78in]{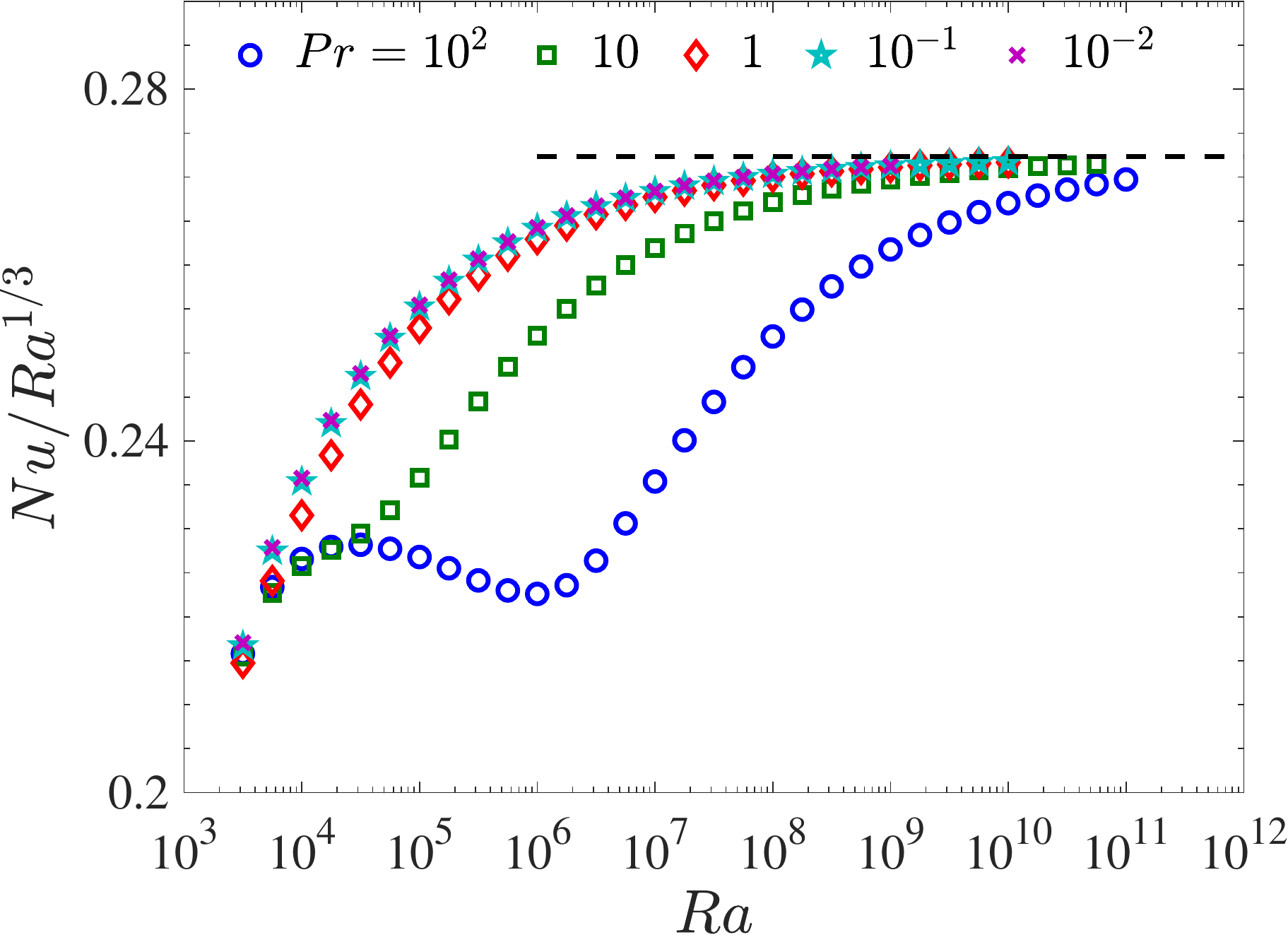}
\end{minipage}\hfill
\begin{minipage}[t]{0.025\textwidth}
\vspace{0pt}
(b)
\end{minipage}
\begin{minipage}[t]{0.46\textwidth}
\vspace{0pt}
\includegraphics[height=1.78in]{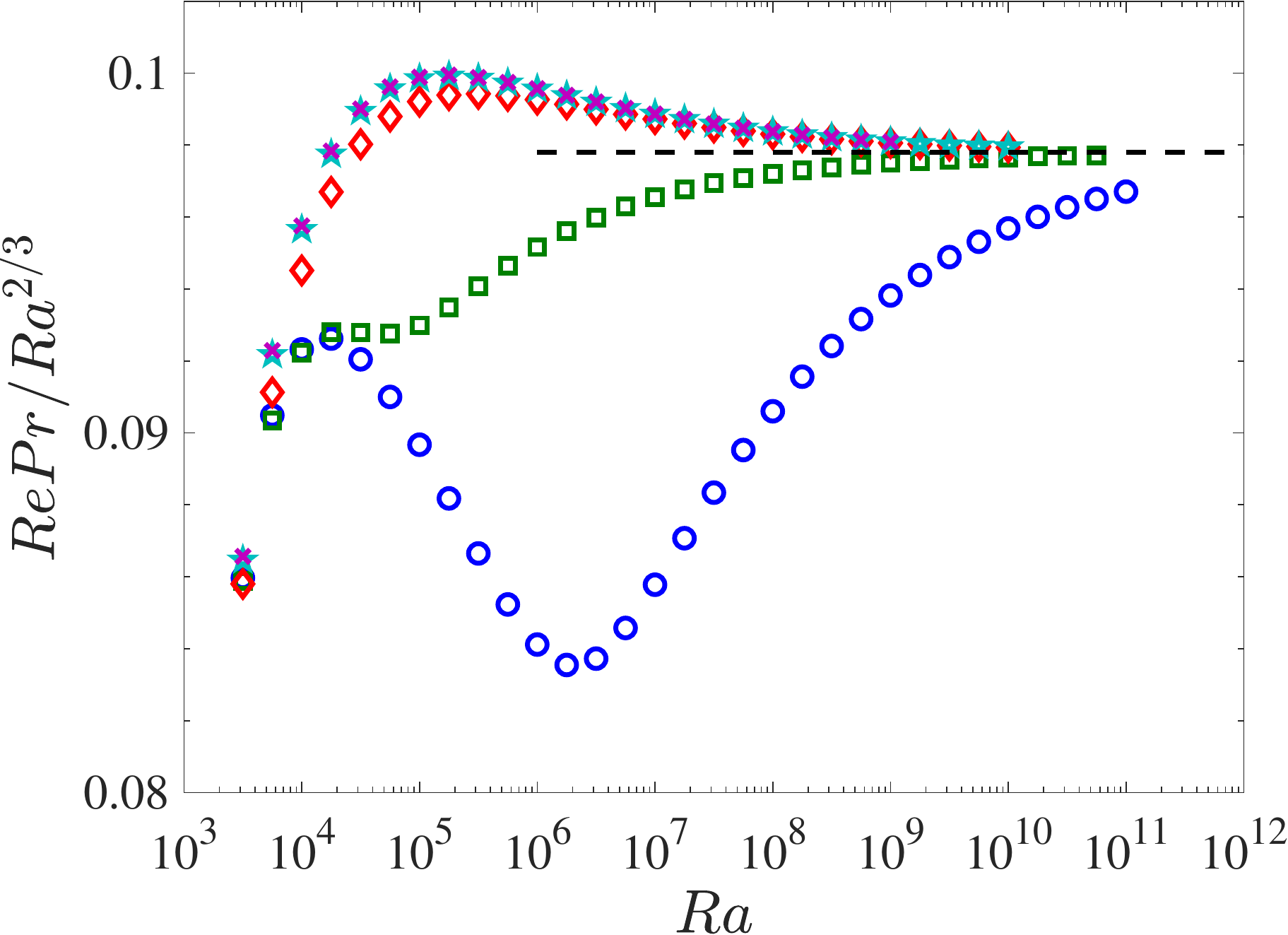}
\end{minipage}
\caption{\label{fig: Ra}The $\Ra$-dependence of (a) $\Nu$ and (b) $\Rey$, compensated by the asymptotic scalings~\cref{eq: scalings}, for steady convective rolls with $\Gamma=2$ ($k=\pi$) at various $\Pran$. Dashed lines in panels (a) and (b) denote, respectively, the asymptotic prefactor $c_n(\pi)\approx0.2723$ from \citet{ChiniCox2009} and our asymptotic prediction $c_r(\pi) \approx 0.0978$. \Cref{fig: Ultimate} shows the same $\Nu$ values \emph{not} compensated by $\Ra^{1/3}$.}
\end{figure}

\Cref{fig: Ra} shows the $\Ra$-dependence of the compensated quantities $\Nu/\Ra^{1/3}$ and $\Rey\Pran/\Ra^{2/3}$ for rolls of aspect ratio $\Gamma=2$ ($k=\pi$) at various $\Pran$. Rolls of this aspect ratio bifurcate from the conduction state at the Rayleigh number $\Ra_c(\pi)=8\pi^4 \approx 779$.  It is clear from \cref{fig: Ra} that both $\Nu$ and the P\'eclet number $\Rey\Pran$ become independent of $\Pran$ as $\Ra \to \infty$, as predicted by the asymptotics of \citet{ChiniCox2009}, and also as $\Ra$ decreases towards the $\Pran$-independent value $\Ra_c$. Convergence to the large-$\Ra$ asymptotic scaling is slower when $\Pran$ is larger, at least over the four decades of $\Pran$ considered here. Numerical values of $\Nu$ and $\Rey$ at large $\Ra$ suggest scaling prefactors that are indistinguishable from the values $c_n(\pi) \approx 0.2723$ and $c_r(\pi) \approx 0.0978$ predicted by asymptotic analysis.

\begin{figure}
\vspace{0.20in}
\begin{minipage}[t]{0.025\textwidth}
\vspace{0pt}
(a)
\end{minipage}
\begin{minipage}[t]{0.46\textwidth}
\vspace{0pt}
\includegraphics[height=1.85in]{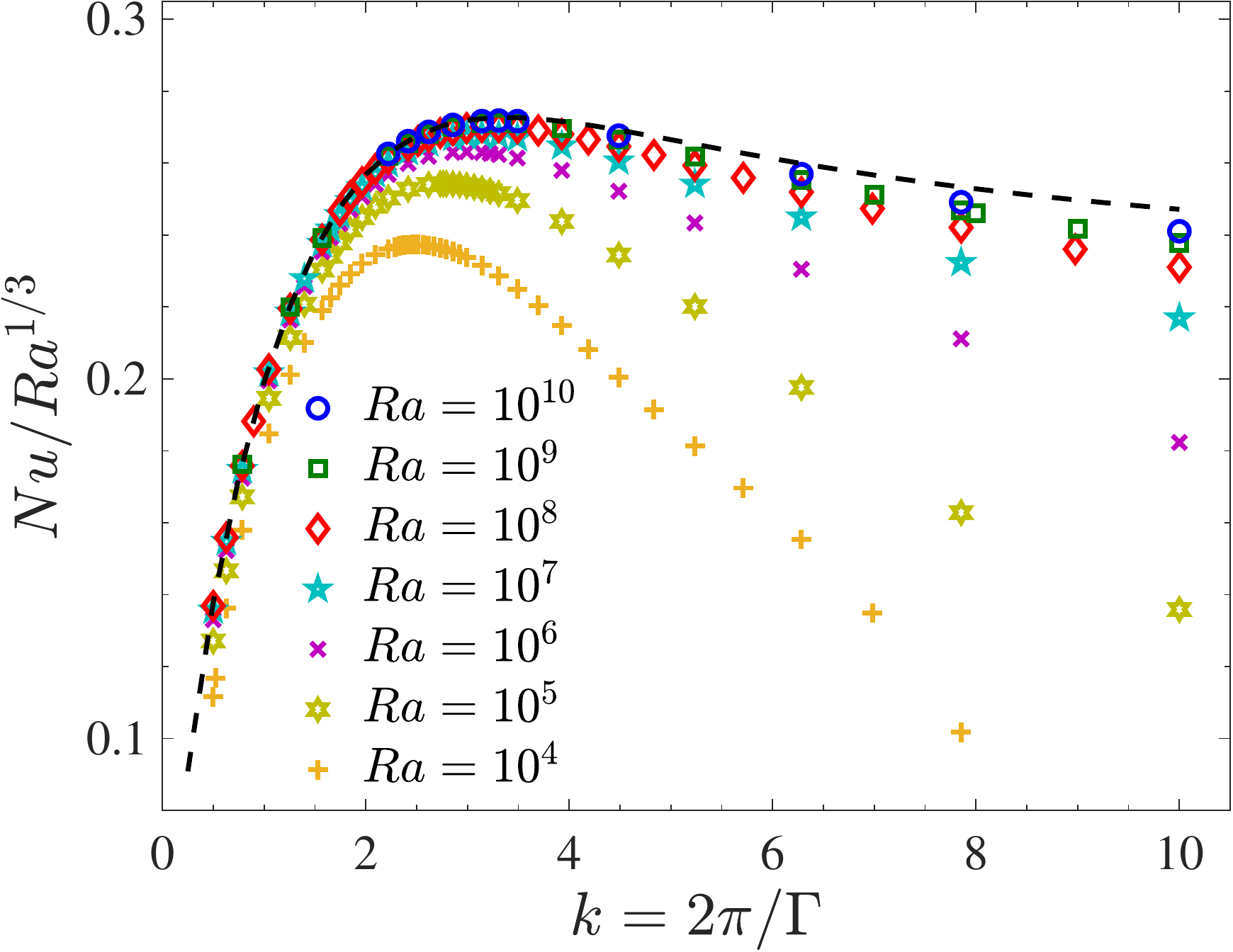}
\end{minipage}\hfill
\begin{minipage}[t]{0.025\textwidth}
\vspace{0pt}
(b)
\end{minipage}
\begin{minipage}[t]{0.46\textwidth}
\vspace{0pt}
\includegraphics[height=1.85in]{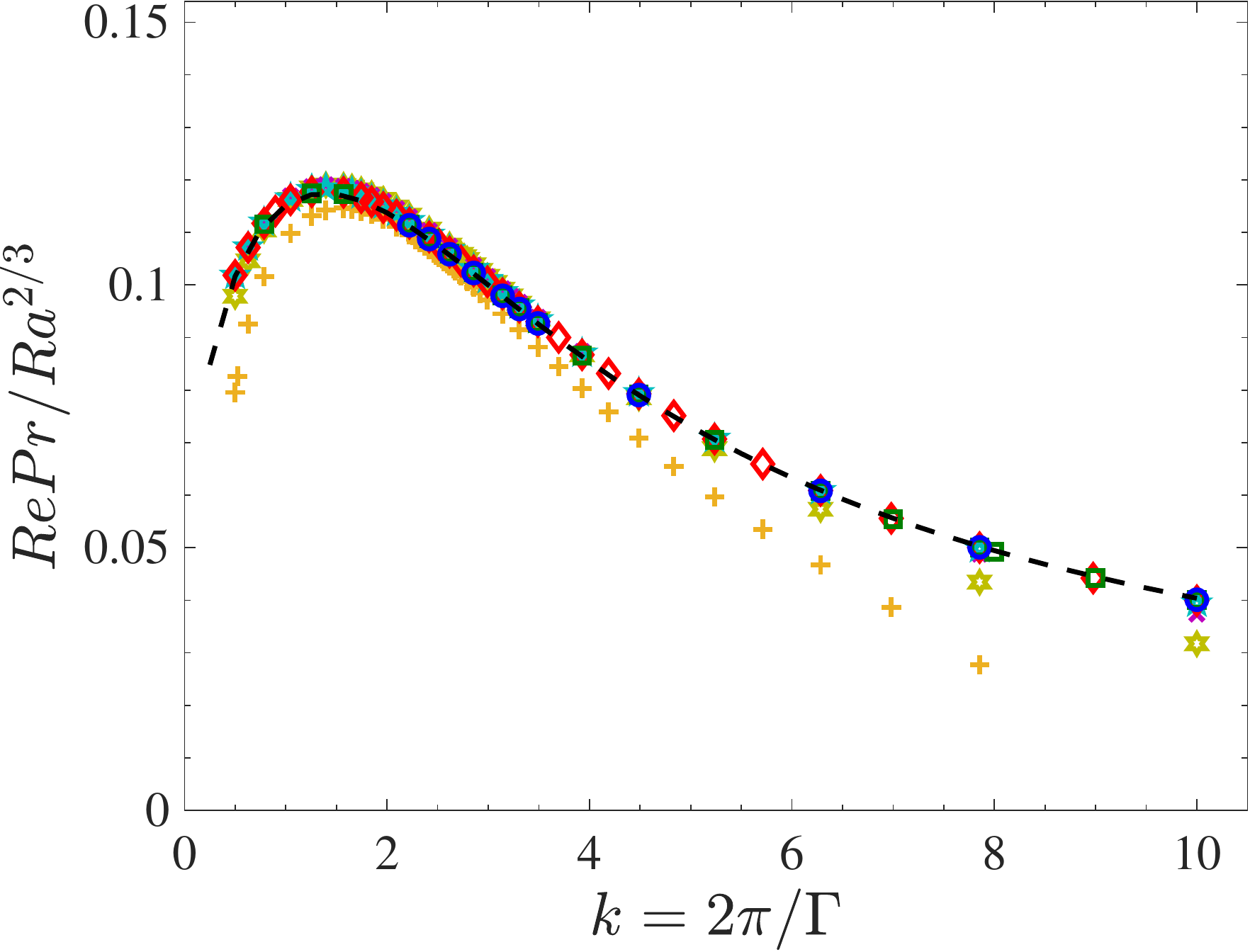}
\end{minipage}
\caption{\label{fig: k}The $k$-dependence of (a) $\Nu$ and (b) $\Rey$, compensated by the asymptotic scalings~\cref{eq: scalings}, for steady convective rolls with $\Pran=1$ at various $\Ra$. The dashed lines in panels (a) and (b) are, respectively, the asymptotic prefactor $c_n(k)$ predicted by \citet{ChiniCox2009} and the corresponding prefactor $c_r(k)$ we derived using their results.}
\end{figure}

\begin{figure}[t!]
\vspace{0.20in}
\begin{minipage}[t]{0.025\textwidth}
\vspace{0pt}
(a)
\end{minipage}
\begin{minipage}[t]{0.46\textwidth}
\vspace{0pt}
\includegraphics[height=1.85in]{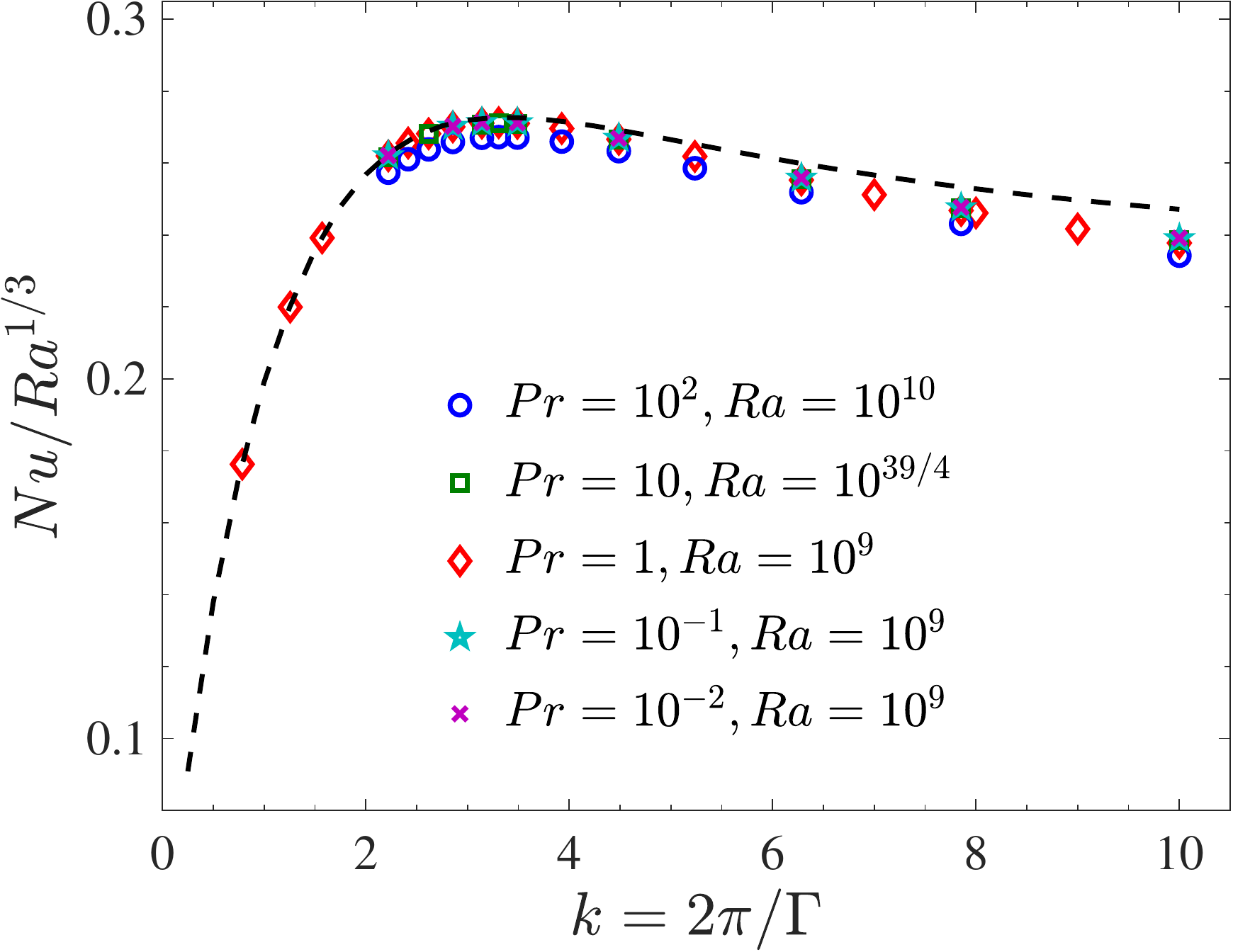}
\end{minipage}\hfill
\begin{minipage}[t]{0.025\textwidth}
\vspace{0pt}
(b)
\end{minipage}
\begin{minipage}[t]{0.46\textwidth}
\vspace{0pt}
\includegraphics[height=1.85in]{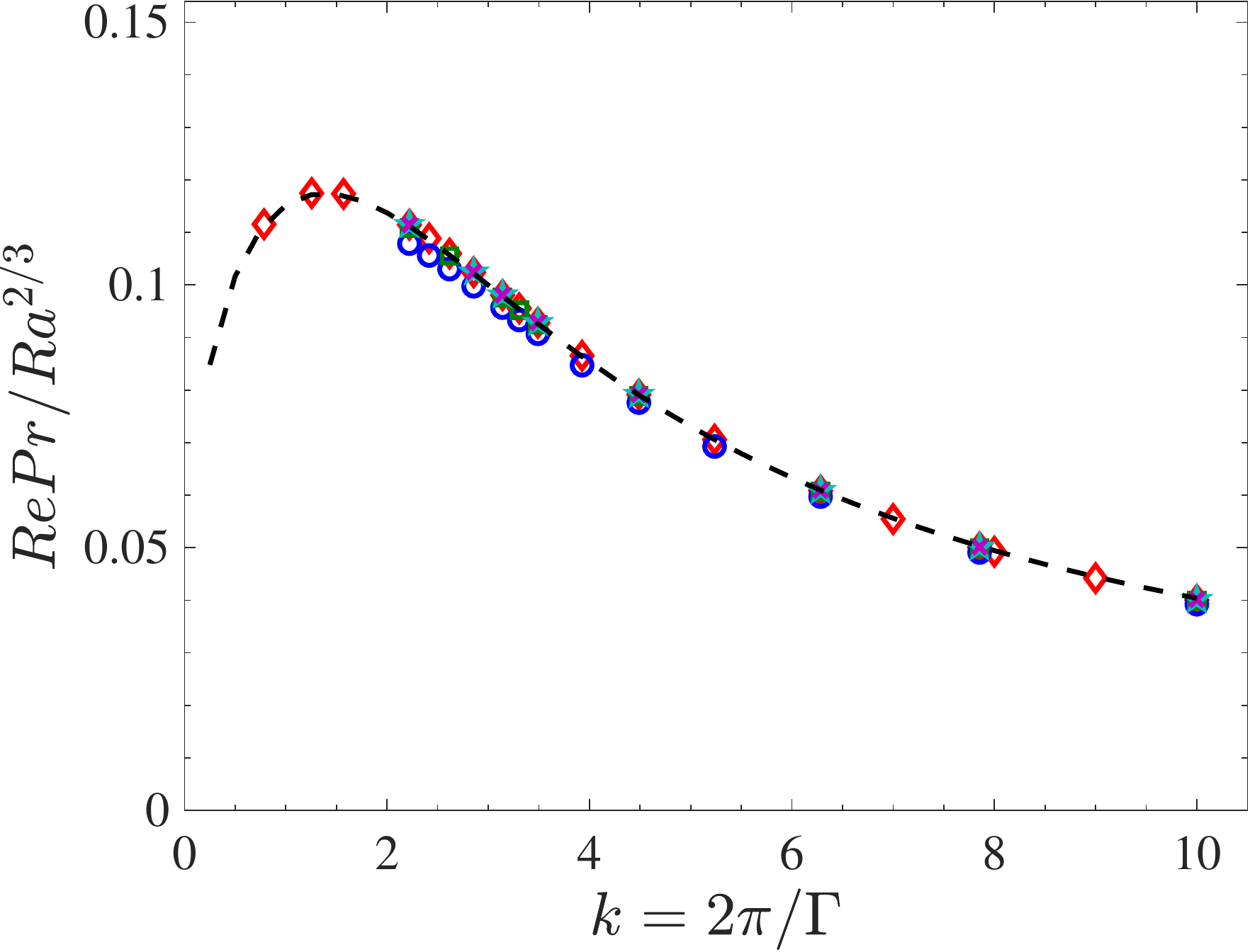}
\end{minipage}
\caption{\label{fig: Pr}Dependence of compensated (a) $\Nu$ and (b) $\Rey$ on $k$ for various $\Pran$ in the large-$\Ra$ asymptotic regime. Reaching this regime requires larger $\Ra$ when $\Pran$ is larger. Asymptotic predictions ($\dashedrule$) of $c_n(k)$ and $c_r(k)$ are as in \cref{fig: k}.}
\end{figure}

\begin{figure}[h!]
\vspace{0.30in}
\begin{minipage}[t]{0.025\textwidth}
\vspace{0pt}
(a)
\end{minipage}
\begin{minipage}[t]{0.46\textwidth}
\vspace{0pt}
\includegraphics[height=1.42in]{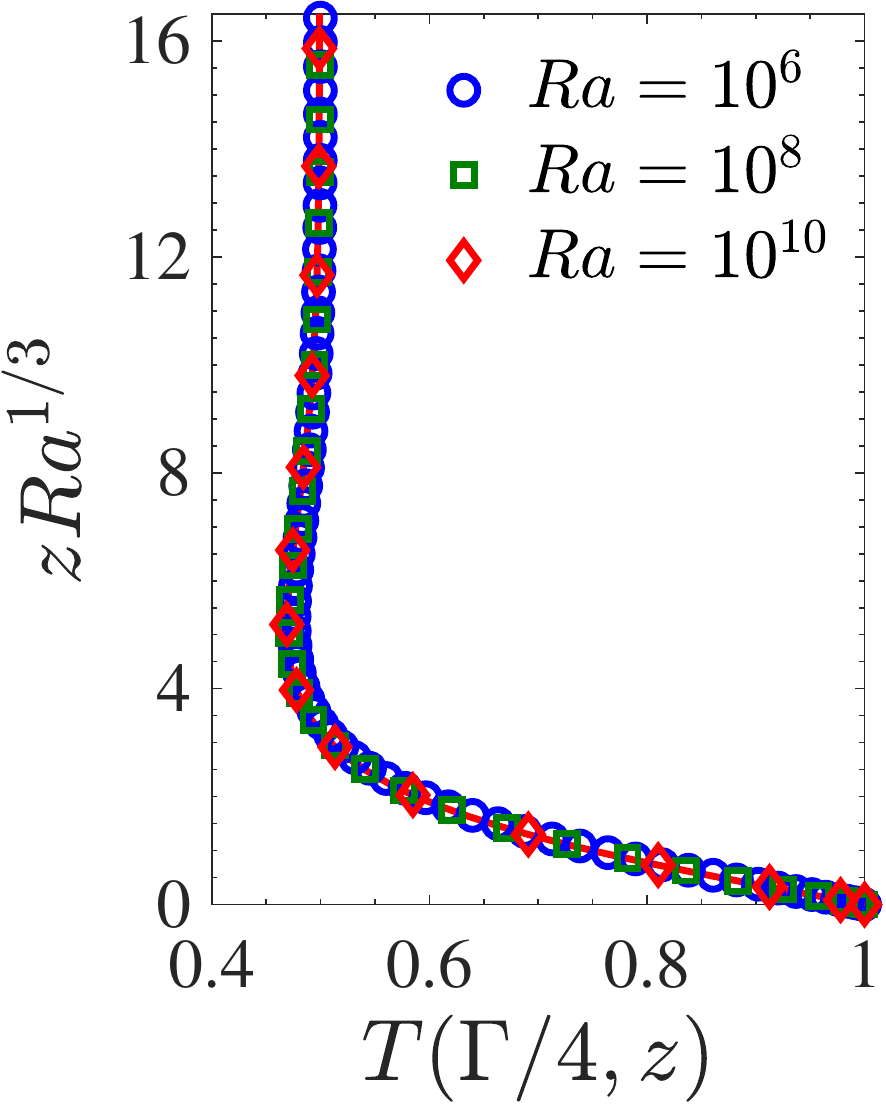} 
\hspace{0.1in}
\includegraphics[height=1.42in]{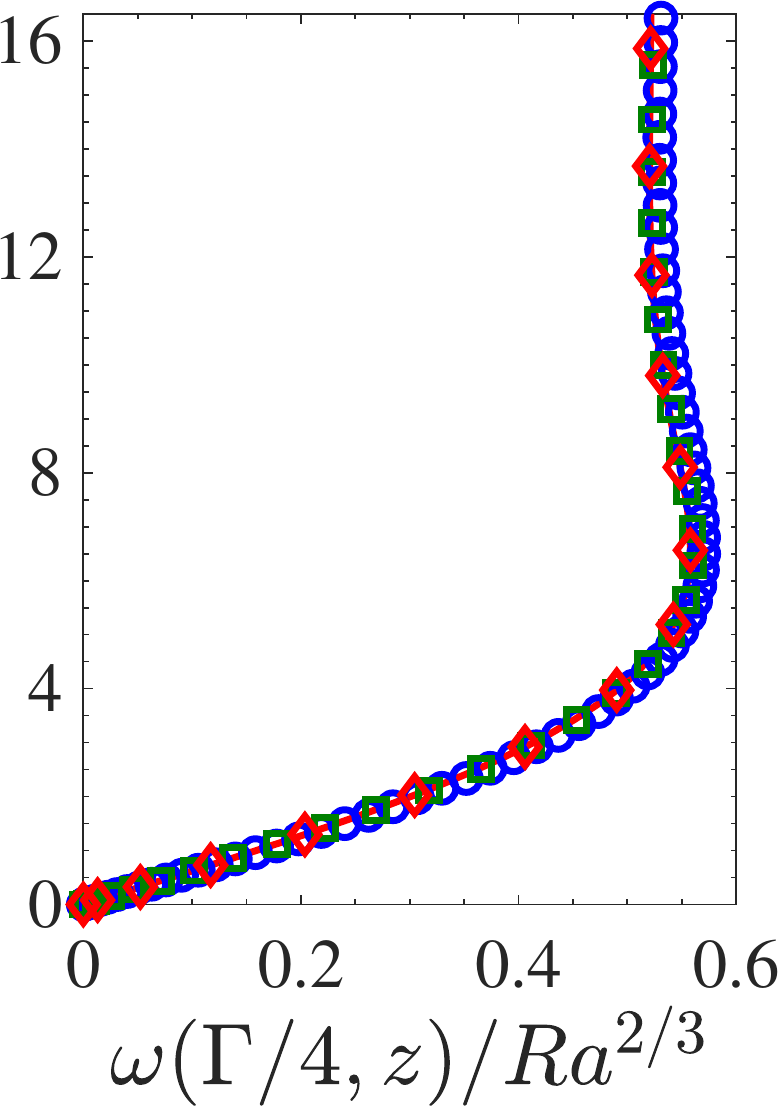}
 \end{minipage}\hfill
\begin{minipage}[t]{0.025\textwidth}
\vspace{0pt}
(b)
\end{minipage}
\begin{minipage}[t]{0.46\textwidth}
\vspace{0pt}
\includegraphics[height=1.42in]{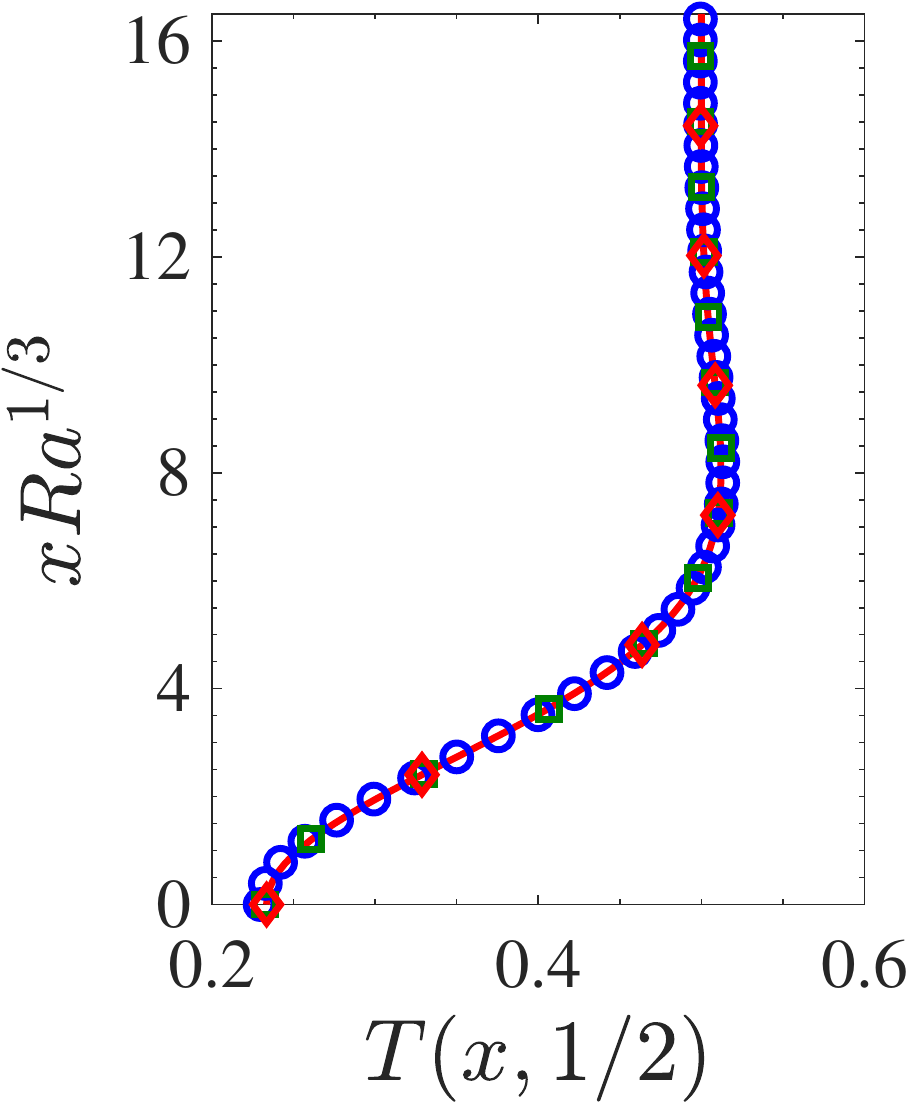} 
\hspace{0.1in}
\includegraphics[height=1.42in]{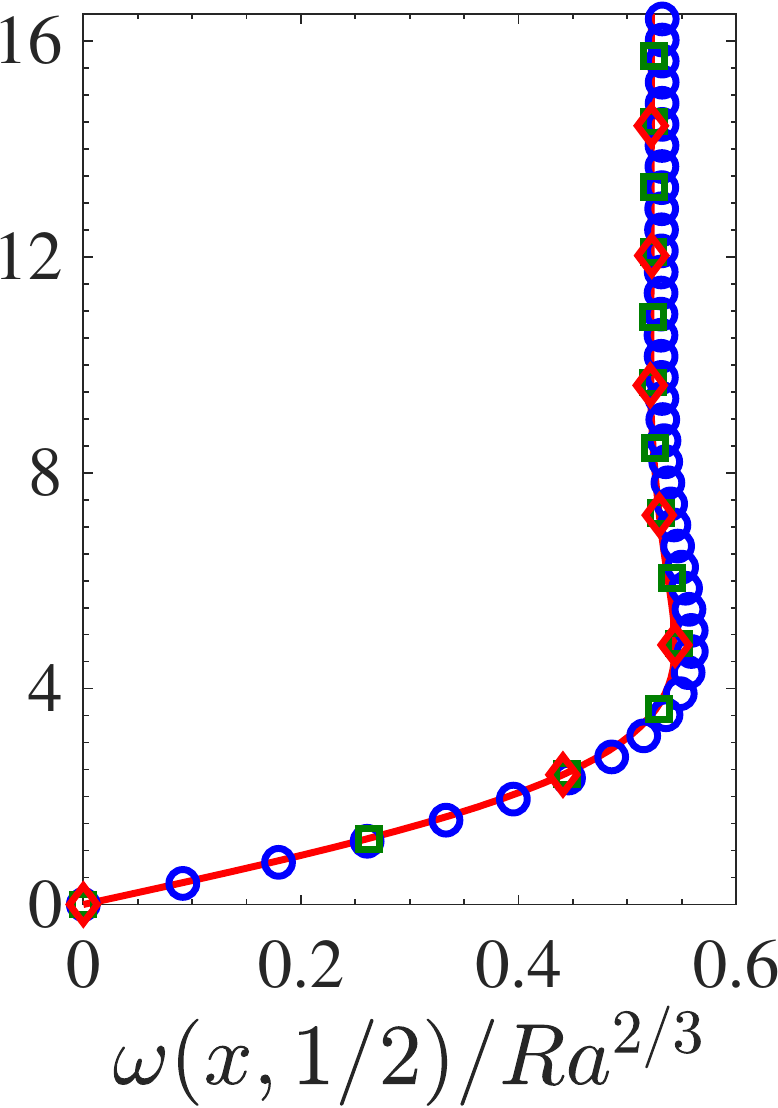}\end{minipage}
\caption{\label{fig: structure} Scaled spatial structure of temperature $T$ and compensated vorticity $\omega$ near the (a) bottom and (b) left side of a convection roll where $\Pran=1$ and $\Gamma=2$. Solid curves are spectral interpolants of $Ra=10^{10}$ values.}
\end{figure}

Nusselt and Reynolds numbers of steady rolls converge to the asymptotic scalings \cref{eq: scalings} over the full range $1/2\le k\le 10$ for which we have computed steady rolls. This is evident in \cref{fig: k} where the $k$-dependence of the compensated quantities $\Nu/\Ra^{1/3}$ and $\Rey \Pran / \Ra^{2/3}$ is shown for various $\Ra$ in the $\Pran=1$ case. As $\Ra$ increases these quantities converge to asymptotic curves that we have called $c_n(k)$ and $c_r(k)$. It is clear from the figure that this convergence is slower when $k$ is larger, and that $\Rey$ reaches its asymptotic scaling sooner than $\Nu$ does. Since rolls with $k=\pi/\sqrt2$ are the first to bifurcate from the conduction state, at $\Ra_c(\pi/\sqrt2)=27\pi^4/4$, this is the $k$ that initially maximizes both $\Nu$ and $\Rey$. As $\Ra\to\infty$, the $k$ values that maximize $\Nu$ and $\Rey$ approach the asymptotic values $k\approx3.31$ ($\Gamma\approx1.9$) and $k\approx1.4$ ($\Gamma\approx4.5$), respectively, where the corresponding maximal prefactors are $c_n\approx0.273$ and $c_r\approx 0.117$.

Both $\Nu$ and the P\'eclet number $\Rey \Pran$ of steady rolls become nearly independent of $\Pran$ as $\Ra$ grows large. The large-$\Ra$ coalescence of data for different $\Pran$ is evident for the $k=\pi$ case in \cref{fig: Ra}, as is the fact that $\Pran$ can have a substantial effect in the pre-asymptotic regime. To show that $\Pran$-independence at large $\Ra$ occurs over the full range $1/2\le k\le 10$ of our computations, \cref{fig: Pr} depicts the $k$-dependence of compensated $\Nu$ and $\Rey$ at large $\Ra$ for various $\Pran$. All $\Nu/\Ra^{1/3}$ and $\Rey\Pran/\Ra^{2/3}$ values plotted in \cref{fig: Pr} fall close to the asymptotic predictions for $c_n(k)$ and $c_r(k)$ that, at leading order in the asymptotic small parameter $\Ra^{-1/3}$, are independent of $\Pran$.

The asymptotic scaling of steady rolls at large $\Ra$ is reflected not only in the collapse of rescaled bulk quantities like $\Nu/\Ra^{1/3}$ and $\Rey \Pran / \Ra^{2/3}$ but also in the collapse of the boundary and internal layer profiles when the appropriate spatial variable is stretched by $\Ra^{1/3}$. \Cref{fig: structure} shows this collapse of the temperature and vorticity profiles at the bottom boundary and at the left edge of the periodic domain for the case where $\Pran=1$ and $\Gamma=2$ ($k=\pi$). Coincidence of these scaled profiles at large $\Ra$ confirms that the thickness of both the thermal and vorticity layers scale like $\Ra^{-1/3}$ on all four edges of a single convection roll, while both fields are strongly homogenized in the interior with $T\sim1/2$ and $\omega \sim 0.522 \Ra^{2/3}$. Profiles at other $\Pran$ are not shown but collapse similarly. These findings confirm the deduction of a homogenized interior by \citet{ChiniCox2009}, as well as their prediction that the core vorticity magnitude is asymptotic to $\sqrt{c_n(\pi)}\Ra^{2/3}\approx 0.5218\Ra^{2/3}$ uniformly in $\Pran$.

Another quantity of interest is the kinetic energy dissipation rate per unit mass,
\begin{align}
\varepsilon(\mathbf{x}^*,t^*) = \frac{\nu}{2} \sum_{i,j=1}^2(\partial_{x_i^*}u_j^* + \partial_{x_j^*}u_i^*)^2,
\label{eq: diss}
\end{align}
where $^*$ denotes dimensional variables. The corresponding bulk viscous dissipation coefficient $C = \langle \varepsilon \rangle h / U^{3}_{rms}$ can be expressed in dimensionless variables as
\begin{align}
C = \Rey^{-3}\Pran^{-2}\,\bigg\langle \frac{1}{2} \sum_{i,j=1}^2(\partial_{i}u_j + \partial_{j}u_i)^2 \bigg\rangle.
\label{eq: diss_nondim}
\end{align}
Identity \cref{eq: NuEquiv} gives $C = \Rey^{-3}\Pran^{-2}\Ra(\Nu-1)$, so the asymptotic scalings~\cref{eq: scalings} imply
\begin{align}
C \sim c_n(k){c_r(k)}^{-3}\Pran\Ra^{-2/3} \sim c_n(k){c_r(k)}^{-2}\Rey^{-1}.
\label{eq: scalings_diss}
\end{align}
That is, $C$ depends asymptotically on $\Ra$ and $\Pran$ via the distinguished combination $\Pran\Ra^{-2/3}$ that is asymptotic to $\Rey^{-1}$. This scaling of the dissipation coefficient is characteristic of flows without viscous boundary layers, such as laminar Couette or Poiseuille flow, consistent with the steady velocity fields computed here (cf.\ \cref{fig: rolls}). Indeed, for stress-free steady convection, viscous dissipation is dominated by that in the homogenized core since the vorticity is of the same asymptotic magnitude in the core as in the thin vorticity layers.

The average of dissipation over time and horizontal directions, denoted $\overline\varepsilon(z)$, has been used to compare convection between the cases of stress-free and no-slip boundaries. In 3D simulations of the stress-free case at $\Ra=5\times 10^6$, \citet{Petschel2013} found that the normalized profile $\overline{\varepsilon}(z)/\langle \varepsilon \rangle$ exhibits `dissipation layers' near the boundaries that depend strongly on $\Pran$.  In steady rolls, on the other hand, we find that $\overline{\varepsilon}(z)/\langle \varepsilon \rangle$ is independent of $\Pran$ at asymptotically large $\Ra$, as shown in figure~S1 of the Supplementary Material.

\section{\label{sec: con}Discussion}

The steady rolls we have computed share many features with unsteady flows from DNS of Rayleigh--B\'enard convection with isothermal stress-free boundary conditions. In recent simulations, \citet{Wang2020} found multistability between unsteady states exhibiting various numbers of roll pairs in wide 2D domains. Each of the coexisting states suggested scalings approaching the $\Nu=\cO(Ra^{1/3})$ and $\Rey=\cO(\Ra^{2/3})$ asymptotic behaviour of steady rolls. As in the steady case, the prefactors of these scalings depended on the mean aspect ratios of the unsteady rolls. The highest Nusselt numbers among \citeauthor{Wang2020}'s data occur in five-roll-pair states in a $\Gamma=16$ domain---meaning each roll pair has $\Gamma\approx3.2$ on average---but steady $\Gamma = 3.2$ rolls have still larger $\Nu$. At $\Ra = 10^9$ and $\Pran=10$, for example, the DNS exhibit $\Nu = 198.01$ and $\Rey = 10135$ while steady $\Gamma = 3.2$ rolls at the same parameters yield the larger values of $\Nu = 253.61$ and $\Rey = 11333$, and comparisons at other $\Ra$ are similar (cf.\ Table~S4 of the Supplementary Material). \Cref{fig: Ultimate} shows the $\Nu$ of these five-roll-pair DNS states along with the larger $\Nu$ of the steady rolls computed here for various $\Pran$ and $\Gamma=2$.  The steady rolls also achieve larger $\Nu$ values than have been attained in other unsteady simulations with stress-free boundaries in 2D \citep{VanderPoel2014a, Goluskin2014} and in 3D \citep{Petschel2013, Pandey2014, Pandey2016, Pandey2016a}.

\begin{figure}
\vspace{0.10in}
{\includegraphics[height=2.35in]{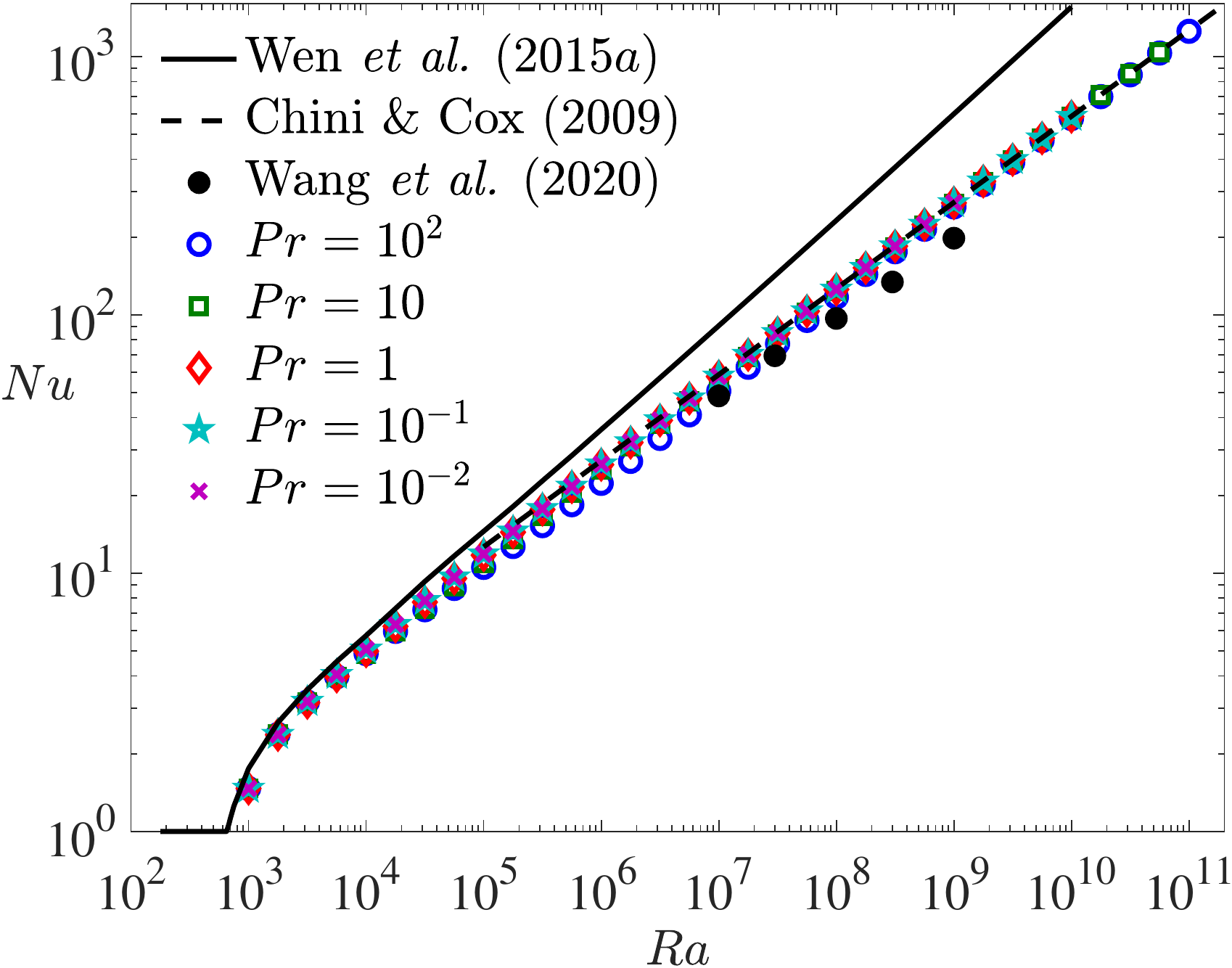}}
\caption{\label{fig: Ultimate}Dependence of $\Nu$ on $\Ra$ for: steady rolls with $\Gamma=2$ and various $\Pran$, time-dependent 2D simulations with mean aspect ratio $\Gamma=3.2$ and $\Pran=10$ \citep[][see text]{Wang2020}, and upper bounds applying to all flows with $\Gamma=2\sqrt2$ and any $\Pran$ \citep[][see text]{Wen2015PRE}. The dashed line is the asymptotic prediction $\Nu \sim 0.2723 \,\Ra^{1/3}$ of \citet{ChiniCox2009} for $\Gamma=2$. The same $\Nu$ values of steady rolls are shown compensated by $\Ra^{1/3}$ in \cref{fig: rolls}.}
\end{figure}

Comparing steady rolls in the stress-free case with those previously computed in the no-slip case, there are significant differences in their dependence on the aspect ratio $\Gamma$. With stress-free boundaries, $\Nu =\cO(Ra^{1/3})$ for each $\Gamma$ as $\Ra\to\infty$, with maximal asymptotic heat transport attained by rolls of optimal aspect ratio $\Gamma\approx1.9$. In the no-slip computations of \citet{Waleffe2015} and \citet{Sondak2015}, on the other hand, the $\Gamma$ values that maximize $\Nu$ decrease towards zero proportionally to $\Ra^{-0.22}$ at large $\Ra$. (A similar phenomenon occurs in porous medium Rayleigh--B\'enard convection; see \citet{Wen2015JFM}.) The no-slip steady rolls display $\Nu$ scaling like $\Ra^{0.28}$ when $\Gamma$ is fixed but scaling like $\Ra^{0.31}$ when the optimal $\Gamma$ is chosen to maximize $\Nu$ at each $\Ra$. The measured exponent 0.31 is unlikely to be exact, so it remains possible that the asymptotic scaling of optimal-$\Gamma$ steady rolls is $\Nu=\cO(\Ra^{1/3})$ in the no-slip case, as in the stress-free case.

Steady rolls in the stress-free and no-slip cases differ also in their dependence on the Prandtl number. Only with stress-free boundaries do the Nusselt number $\Nu$ and P\'eclet number $\Rey\Pran$ apparently become uniform in $\Pran$ as $\Ra\to\infty$. To see how this $Pr$-independence emerges in the stress-free scenario, first note that area-integrated work by the buoyancy forces must balance area-integrated viscous dissipation---i.e., $\Ra \lan wT\ran=\lan |\bnabla u|^2 + |\bnabla w|^2 \ran$ in the dimensionless formulation of \cref{eq: Nu} and \cref{eq: NuEquiv}. The former integral is dominated by plumes since the roll's core is isothermal, so it scales proportionally to the dimensional quantity $\alpha \Delta g \delta hU_{rms} $, where $\delta$ is the dimensional plume thickness. The latter integral is dominated by the core since the vorticity is $\cO(U_{rms}/h)$ everywhere in the stress-free case, so this integral scales proportionally to the dimensional quantity $\nu (U_{rms}/h)^2 h^2$. Balancing advection with diffusion of temperature anomalies in the thermal boundary layers, which also have thickness $\delta$, requires that $U_{rms}$ scale in proportion to $\kappa h/\delta^2$.  Combining this scaling relationship with that from the integral balance gives the dimensionless thermal boundary layer thickness as $\delta/h = \cO(\Ra^{-1/3})$---and so $\Nu = \cO(Ra^{-1/3})$---uniformly in $\Pran$.  These relationships also imply that $U_{rms}$ is proportional to $(\kappa/h) \Ra^{2/3}$, and so the P\'eclet number $\Rey\Pran=U_{rms}h/\kappa$ scales as $\Ra^{2/3}$ uniformly in $\Pran$.  Ultimately, it is the passivity of the vorticity boundary layers that results in the $Pr$-independence of these emergent bulk quantities.  The vorticity layers not only make no contribution to the total dissipation at leading order, they also have no leading-order effect on the stream function that is responsible for the convective flux $\langle wT\rangle$.

The $\Rey=\cO(\Ra^{2/3})$ scaling found at large $\Ra$ for steady rolls and approximately evidenced in the DNS of \citet{Wang2020} means that buoyancy forces can sustain substantially faster-than-free-fall velocities. Indeed, if flow speeds were limited by the maximum buoyancy acceleration acting across the layer height then dimensional characteristic velocities could not be of larger order than $\sqrt{g \alpha \Delta h}$, and $\Rey$ could not be larger than $\cO(\Ra^{1/2})$. Such $\Rey$ may be expected if the bulk flow were dominated by effectively independent rising and falling plumes. Significantly higher speeds apparently persist within coherent convection rolls, whether steady or unsteady.

Although steady rolls cannot give heat transport larger than $\Nu=\cO(\Ra^{1/3})$ as \mbox{$\Ra\to\infty$} in the stress-free case \citep{ChiniCox2009}, it is an open question whether larger $\Nu$ can result from time-dependent flows or other types of steady states in either 2D or 3D. Rigorous upper bounds on $\Nu$ derived from the governing equations---bounds depending on $\Ra$ that apply to all flows regardless of whether they are steady or unsteady and stable or unstable---do not rule out $\Nu$ growing faster than $\Ra^{1/3}$. Specifically, for the 2D stress-free case \citet{Whitehead2011} proved analytically that $\Nu \le 0.289 \, \Ra^{5/12}$ uniformly in both $\Pran$ and domain aspect ratio. \Citet{Wen2015PRE} improved the prefactor of this bound by solving the relevant variational problem numerically, computing bounds up to large finite $\Ra$ with a prefactor depending weakly on $\Gamma$. The numerical upper bound they computed for $\Gamma=2\sqrt2$ is shown in \cref{fig: Ultimate}; its scaling at large $\Ra$ is $\Nu \le 0.106\, \Ra^{5/12}$. For 3D flows in the stress-free case only the larger upper bound $\Nu\le\cO(\Ra^{1/2})$ has been proved \citep{Doering1996}. It remains to be seen whether upper bounds smaller than $\cO(\Ra^{5/12})$ or $\cO(\Ra^{1/2})$ can be proved for 2D or 3D flows, respectively, and whether there exists any sequence of solutions for which $\Nu$ grows faster than $\cO{(\Ra^{1/3})}$. In view of available evidence it is possible that, at each $\Ra$ and $\Pran$, the steady 2D roll with the largest value of $\Nu$---i.e., with $\Nu$ maximized over $\Gamma$---transports more heat than any other 2D or 3D solution. We are aware of no counterexamples to this possibility, either in the stress-free case studied here or in the no-slip case.

\section*{Acknowledgements}
\addcontentsline{toc}{section}{\protect\numberline{}Acknowledgements}
We thank L.M.\ Smith, D.\ Sondak and F.\ Waleffe for helpful discussions.
This research was supported in part by US National Science Foundation awards DMS-1515161 and DMS-1813003, Canadian NSERC Discovery Grants Program awards RGPIN-2018-04263, RGPAS-2018-522657 and DGECR-2018-00371, and computational resources and services provided by Advanced Research Computing at the University of Michigan.

\section*{Declaration of interests}
\addcontentsline{toc}{section}{\protect\numberline{}Declaration of interests}
The authors report no conflict of interest.

\appendix
\section{\label{appA}Asymptotic calculation of $RePr/Ra^{2/3}$}

In this Appendix we derive the asymptotic scaling relation for $\Rey$ that follows from the asymptotic analysis of \citet{ChiniCox2009} but is not reported there. In the large-$\Ra$ limit, a steady roll's velocity field is properly scaled by $(\kappa/h)Ra^{2/3}$ rather than by the thermal diffusion velocity $\kappa/h$.  Accordingly, we introduce the rescaled dimensionless velocity $(u_\infty,w_\infty)$, which is related to the dimensionless velocity in \cref{eq: bouss} by $(u,w) = Ra^{2/3}(u_\infty,w_\infty)$.  With this rescaling, the expression \cref{eq: Re} for $\Rey$ becomes
\beq
Re=\frac{1}{Pr}\left\langle u_\infty^2 + w_\infty^2 \right\rangle^{1/2} Ra^{2/3}\, = \,\frac{1}{Pr}\left\langle |\nabla \psi_\infty|^2 \right\rangle^{1/2} Ra^{2/3},\label{ReRaAsymptotic}
\eeq
where $\psi_\infty$ is the correspondingly rescaled stream function. Consequently, the prefactor in the asymptotic relation \cref{eq: scalings} for $\Rey$ satisfies
\begin{align}
c_r= \left\langle |\nabla \psi_\infty|^2 \right\rangle^{1/2}. 
\label{CrAsymptotic}
\end{align}
To evaluate the right-hand side of \eqref{CrAsymptotic} in $Ra\to\infty$ limit we first integrate by parts to find
\begin{align}
\int_0^1\int_0^{\pi/k} |\nabla\psi_\infty|^2\mathrm{d}x\mathrm{d}z=\left|\int_0^1\int_0^{\pi/k} \psi_\infty \omega_\infty \mathrm{d}x\mathrm{d}z\right|\,\sim\,\Omega_c(k)\left|\int_0^1\int_0^{\pi/k} \psi_\infty(x,z) \mathrm{d}x\mathrm{d}z\right|,
\label{IntegByParts}
\end{align}
where $\nabla^2\psi_\infty=-\omega_\infty$. The asymptotic approximation in (\ref{IntegByParts}) follows from two asymptotic estimates. First, the vorticity $\omega_\infty(x,z)$ in a steady roll's core is homogenized to a spatially uniform value $\Omega_c$ as $Ra\to\infty$, and according to the analysis of \citeauthor{ChiniCox2009} this value is related the prefactor in the $Nu$--$Ra$ asymptotic relation via $\Omega_c\sim\sqrt{c_n(k)}$. Second, owing to the stress-free conditions and symmetries on a steady roll's periphery, the vorticity field is of the same asymptotic order in both the vorticity boundary layers and the core, so the contribution of these boundary layers to the middle integral in \eqref{IntegByParts} is negligible relative to that from the $\mathit{O}(1)$ core where $\omega_\infty\sim\Omega_c$.

Unlike the temperature and vorticity fields, the stream function $\psi_\infty$ at leading order in $Ra$ is a smooth function over the entire spatial domain. The leading order expression for $\psi_\infty$ given by equations (22)--(23) of \citeauthor{ChiniCox2009} is, in our notation,
\begin{align}
\psi_\infty(x,z)\sim\sum_{n=1,\,\mathrm{odd}}^{\infty}\psi_n(z)\sin{(nkx)}\,=\,
   \sum_{n=1,\,\mathrm{odd}}^{\infty}\frac{4\Omega_c}{\pi k^2 n^3}\left[1-\frac{\cosh{(nk(z-1/2))}}{\cosh{(nk/2)}}\right]\sin{(nkx)}.
\label{eq: psi inf}
\end{align}
Substituting \cref{eq: psi inf} into the right-hand side of \cref{IntegByParts}, integrating term-by-term, dividing by the cell width $\pi/k$ to obtain the spatial average, and taking the square root of the resulting expression gives the asymptotic prediction
\beq
c_r(k)\sim\left(\frac{8c_n(k)}{\pi^2k^2}\,\sum_{n=1,\,\mathrm{odd}}^{\infty}\frac{1}{n^4}\left[1-\frac{2\tanh{(nk/2)}}{nk}\right]\right)^{1/2}.
\eeq
Values of $c_n(k)$ and corresponding values of $c_r(k)$ for various $k\in[1/4,10]$ are given in Table S1 of the Supplementary Material.

\newpage

\section*{Supplementary Material}
\addcontentsline{toc}{section}{\protect\numberline{}Supplemental Material}

Table~S1 gives numerical values of the asymptotic prefactors in expression (3.1) for the steady rolls constructed in the asymptotic analysis of \cite{ChiniCox2009}. Tables~S2 and~S3 give the $\Nu$ and $\Rey\Pran$ values plotted in figure 2. Table~S4 compares steady rolls to unsteady rolls with the same mean aspect ratio from the DNS of \cite{Wang2020}.

Figure~S1 shows the normalized dissipation profiles $\overline{\varepsilon}(z)/\langle \varepsilon \rangle$ of steady rolls at various $\Pran$ for $\Ra$ values of $5\times 10^6$ and $10^9$. There is much less variation with $\Pran$ at the larger $\Ra$ value, reflecting the $\Pran$-independence that is predicted in the $\Ra\to\infty$ asymptotic limit. Only the $\Pran=100$ profile is significantly different from the others when $\Ra=10^9$; convergence to asymptotic behaviour as $\Ra\to\infty$ is slower at larger $\Pran$ (cf.\ figure 1).

\begin{table}[h!]
 \begin{center}
\begin{tabular}{ccccccc}
\hline
$k=2\pi/\Gamma$	&	$c_n(k)$	  &	$c_r(k)$	&  \quad\quad\quad &	 $k=2\pi/\Gamma$	&	$c_n(k)$	  &	$c_r(k)$\\ 
\hline	
0.25	&	0.09084366	&	0.08479653	& &	5.25	&	0.26516638	&	0.07042333	\\
0.5	&	0.13783563	&	0.10165665	& &	5.5	&	0.26383550	&	0.06791123	\\
0.75	&	0.17242486	&	0.11048428	& &	5.75	&	0.26253039	&	0.06554476	\\
1	&	0.19909232	&	0.11516486	& &	6	&	0.26126222	&	0.06331554	\\
1.25	&	0.21981992	&	0.11716681	& &	6.25	&	0.26003730	&	0.06121496	\\
1.5	&	0.23579393	&	0.11727251	& &	6.5	&	0.25886080	&	0.05923472	\\
1.75	&	0.24788488	&	0.11600002	& &	6.75	&	0.25773340	&	0.05736652	\\
2	&	0.25681088	&	0.11373790	& &	7	&	0.25665667	&	0.05560274	\\
2.25	&	0.26318917	&	0.11079160	& &	7.25	&	0.25563005	&	0.05393607	\\
2.5	&	0.26754979	&	0.10740171	& &	7.5	&	0.25465196	&	0.05235964	\\
2.75	&	0.27033921	&	0.10375413	& &	7.75	&	0.25372157	&	0.05086718	\\
3	&	0.27192661	&	0.09998906	& &	8	&	0.25283657	&	0.04945277	\\
3.25	&	0.27260954	&	0.09620876	& &	8.25	&	0.25199552	&	0.04811103	\\
3.5	&	0.27262477	&	0.09248559	& &	8.5	&	0.25119483	&	0.04683679	\\
3.75	&	0.27215669	&	0.08886847	& &	8.75	&	0.25043350	&	0.04562556	\\
4	&	0.27134834	&	0.08538889	& &	9	&	0.24970964	&	0.04447307	\\
4.25	&	0.27030747	&	0.08206516	& &	9.25	&	0.24902004	&	0.04337532	\\
4.5	&	0.26911578	&	0.07890636	& &	9.5	&	0.24836456	&	0.04232885	\\
4.75	&	0.26783386	&	0.07591494	& &	9.75	&	0.24773960	&	0.04133018	\\
5	&	0.26650668	&	0.07308893	& &	10	&	0.24714363	&	0.04037628	\\
\hline
  \end{tabular}
  \caption*{Table S1: Numerical values of the asymptotic prefactors $c_n$ and $c_r$ in (3.1) for various wavenumbers $k$. Values of $c_n$ are from the data of \citet{ChiniCox2009}, and $c_r$ is calculated from $c_n$ according to \mbox{(A\,5)}.}{\label{tab:cncr}}
 \end{center}
\end{table}

\vspace*{\fill}
\begin{table}[h!]
 \begin{center}
\begin{tabular}{ccccccc}
\hline
\multirow{2}{*}{$Ra$} &	\multirow{2}{*}{$N_x \times N_z$}			&			&			&  $Nu$	&		&  \\ 
 &  & $Pr=10^{-2}$ & $Pr=10^{-1}$ & $Pr=1$ & $Pr=10$ & $Pr=10^2$ \\ \hline
$10^3$	 & 		$128 \times 65$	 & 1.46630	     & 1.46614	& 1.46687		& 1.46716		&1.46718\\
$10^{{13}/{4}}$ &	$128 \times 65$ & 2.37255 & 2.37025 & 2.36637 & 2.37324	 & 2.37425\\
$10^{{14}/{4}}$ & $128 \times 65$ & 3.18564 & 3.18049 & 3.15193 & 3.16265 & 3.16748\\
$10^{{15}/{4}}$ & $128 \times 65$ & 4.05203 & 4.04468 & 3.98471 & 3.96052 & 3.97220\\
$10^4$ & $128 \times 65$ & 5.07914 & 5.07051 & 4.98831 & 4.86435 & 4.88129\\
$10^{{17}/{4}}$ & $128 \times 65$ & 6.32515 & 6.31587 & 6.22172 & 5.94155 & 5.94945\\
$10^{{18}/{4}}$ & $128 \times 65$ & 7.83124 & 7.82158 & 7.72035 & 7.25591 & 7.21592\\
$10^{{19}/{4}}$ & $128 \times 65$ & 9.65227 & 9.64234 & 9.53590 & 8.89159 & 8.72489\\
$10^5$ & $128 \times 65$ & 11.8568 & 11.8467 & 11.7360 & 10.9457 & 10.5267\\
$10^{{21}/{4}}$ & $256 \times 97$ & 14.5268 & 14.5164 & 14.4021 & 13.5059 & 12.6816\\
$10^{{22}/{4}}$ & $256 \times 97$ & 17.7612 & 17.7506 & 17.6329 & 16.6557 & 15.2701\\
$10^{{23}/{4}}$ & $256 \times 97$ & 21.6798 & 21.6690 & 21.5483 & 20.5044 & 18.4060\\
$10^6$ & $256 \times 97$ & 26.4274 & 26.4164 & 26.2929 & 25.1935 & 22.2598\\
$10^{{25}/{4}}$ & $256 \times 97$ & 32.1797 & 32.1685 & 32.0425 & 30.8962 & 27.0879\\
$10^{{26}/{4}}$ & $256 \times 97$ & 39.1494 & 39.1379 & 39.0094 & 37.8234 & 33.2279\\
$10^{{27}/{4}}$ & $512 \times 129$ & 47.5938 & 47.5822 & 47.4514 & 46.2312 & 41.0104\\
$10^7$ & $512 \times 129$ & 57.8250 & 57.8132 & 57.6802 & 56.4307 & 50.7115\\
$10^{{29}/{4}}$ & $512 \times 129$ & 70.2211 & 70.2091 & 70.0740 & 68.7989 & 62.6608\\
$10^{{30}/{4}}$ & $512 \times 129$ & 85.2397 & 85.2277 & 85.0905 & 83.7928 & 77.2962\\
$10^{{31}/{4}}$ & $512 \times 129$ & 103.437 & 103.424 & 103.285 & 101.967 & 95.1593\\
$10^8$ & $512 \times 129$ & 125.484 & 125.472 & 125.330 & 123.991 & 116.909\\
$10^{{33}/{4}}$ & $768 \times 193$ & 152.192 & 152.179 & 152.036 & 150.683 & 143.355\\
$10^{{34}/{4}}$ & $1024 \times 193$ & 184.554 & 184.541 & 184.394 & 183.026 & 175.477\\
$10^{{35}/{4}}$ & $1024 \times 257$ & 223.758 & 223.745 & 223.599 & 222.215 & 214.465\\
$10^9$ & $1024 \times 257$ & 271.266 & 271.253 & 271.105 & 269.698 & 261.773\\
$10^{{37}/{4}}$ & $1280 \times 257$ &   & 328.798 & 328.713 & 327.238 & 319.134\\
$10^{{38}/{4}}$ & $1280 \times 257$ &   & 398.523 & 398.427 & 397.003 & 388.734\\
$10^{{39}/{4}}$ & $1536 \times 257$ &   & 483.062 & 482.910 & 481.470 & 473.049\\
$10^{10}$ & $1792 \times 257$       &  & 585.437 & 585.285 & 583.841 & 575.311\\
$10^{{41}/{4}}$ & $2048 \times 321$ &  &	&  & 707.958 & 699.254\\
$10^{{42}/{4}}$ & $2560 \times 321$ &  &    &  & 858.111 & 849.325\\
$10^{{43}/{4}}$ & $3072\times 321$  &  &    &  & 1040.21 & 1031.35\\
$10^{11}$ &  $3584 \times 321$	 &  &    &  &  & 1251.98\\
\hline
  \end{tabular}
  \caption*{\label{tab:fig2a}Table S2: Values of the Nusselt number ($\Nu$) plotted in figure~2($a$) for steady rolls of aspect ratio $\Gamma=2$ ($k=\pi$). The resolution of Fourier modes ($N_x$) and Chebyshev collocation points ($N_z$) is given also.}
 \end{center}
\end{table}
\vspace*{\fill}

\newpage

\vspace*{\fill}
\begin{table}[h!]
 \begin{center}
\begin{tabular}{ccccccc}
\hline
\multirow{2}{*}{$Ra$} &	\multirow{2}{*}{$N_x \times N_z$}			&			&			&  $Re Pr$	&		&  \\ 
 &  & $Pr=10^{-2}$ & $Pr=10^{-1}$ & $Pr=1$ & $Pr=10$ & $Pr=10^2$ \\ \hline
$10^3$	 & 		$128 \times 65$ & 4.860211   &  4.859365   &  4.863013   &  4.864534   &  4.864669\\
$10^{{13}/{4}}$ &	$128 \times 65$ & 11.11289   &  11.10342   &  11.08274   &  11.10817   &  11.11225\\
$10^{{14}/{4}}$ & $128 \times 65$ & 18.64998   &  18.62765   &  18.48547   &  18.50202   &  18.52133\\
$10^{{15}/{4}}$ & $128 \times 65$ & 29.18438   &  29.14861   &  28.81749   &  28.56985   &  28.61439\\
$10^4$ & $128 \times 65$ & 44.44510   &  44.39718   &  43.87005   &  42.81088   &  42.85149\\
$10^{{17}/{4}}$ & $128 \times 65$ & 66.65392   &  66.59490   &  65.87916   &  63.22068   &  63.10225\\
$10^{{18}/{4}}$ & $128 \times 65$ & 99.00479   &  98.93476   &  98.02324   &  92.79036   &  92.04534\\
$10^{{19}/{4}}$ & $128 \times 65$ & 146.2211   &  146.1395   &  145.0123   &  136.1512   &  133.5699\\
$10^5$ & $128 \times 65$ & 215.2057   &  215.1114   &  213.7389   &  200.3372   &  193.1899\\
$10^{{21}/{4}}$ & $256 \times 97$ & 316.0649   &  315.9563   &  314.2983   &  295.6302   &  278.8445\\
$10^{{22}/{4}}$ & $256 \times 97$ & 463.6093   &  463.4841   &  461.4878   &  436.6766   &  402.1873\\
$10^{{23}/{4}}$ & $256 \times 97$ & 679.5490   &  679.4041   &  677.0022   &  644.8536   &  580.6680\\
$10^6$ & $256 \times 97$ & 995.7158   &  995.5474   &  992.6546   &  951.6627   &  841.1621\\
$10^{{25}/{4}}$ & $256 \times 97$ & 1458.790   &  1458.594   &  1455.104   &  1403.377   &  1226.322\\
$10^{{26}/{4}}$ & $256 \times 97$ & 2137.241   &  2137.009   &  2132.789   &  2067.960   &  1803.793\\
$10^{{27}/{4}}$ & $512 \times 129$ & 3131.499	   &  3131.224   &  3126.109   &  3045.229   &  2674.566\\
$10^7$ & $512 \times 129$ & 4588.895   &  4588.574   &  4582.361   &  4481.765   &  3981.512\\
$10^{{29}/{4}}$ & $512 \times 129$ & 6725.606   &  	6725.221   &  6717.660   &  6592.805   &  5932.137\\
$10^{{30}/{4}}$ & $512 \times 129$ & 9858.786	   &  9858.334   &  9849.112   &  9694.374   &  8833.901\\
$10^{{31}/{4}}$ & $512 \times 129$ & 14453.85	 &  14453.31   &  14442.05   &  14250.46   &  13140.24\\
$10^8$ & $512 \times 129$ & 21193.84   &  21193.21   &  21179.44   &  20942.25   &  19519.46\\
$10^{{33}/{4}}$ & $768 \times 193$ & 31080.45	   &  31079.68   &  31062.85   &  30769.81 &  28954.88\\
$10^{{34}/{4}}$ & $1024 \times 193$ & 45585.16   &  45584.26   &  45563.33   &  45201.19   &  42895.45\\
$10^{{35}/{4}}$ & $1024 \times 257$ & 66865.43   &  66864.32   &  66839.10   &  66391.72   &  63471.97\\
$10^9$ & $1024 \times 257$ & 98091.13   &  98089.85   &  98058.98   &  97504.78   &  93819.57\\
$10^{{37}/{4}}$ & $1280 \times 257$ &   &  143906.3   &  143882.5   &  143186.5   &  138542.0\\
$10^{{38}/{4}}$ & $1280 \times 257$ &   &  211139.3   &  211107.6   &  210266.2   &  204428.0\\
$10^{{39}/{4}}$ & $1536 \times 257$ &   &  309825.2   &  309769.0   &  308731.1   &  301407.4\\
$10^{10}$ & $1792 \times 257$       &   &  454641.9   &  454573.4   &  453293.9   &  444122.9\\
$10^{{41}/{4}}$ & $2048 \times 321$ &   &    &    &  665545.3   &  654078.8\\
$10^{{42}/{4}}$ & $2560 \times 321$ &   &    &     &  977028.2   &  962715.7\\
$10^{{43}/{4}}$ & $3072\times 321$  &   &     &     &  1434327   &  1416484\\
$10^{11}$ &  $3584 \times 321$	&   &   &   &   &  2083475\\
\hline
  \end{tabular}
    \caption*{\label{tab:fig2b}Table S3: Values of the P\'eclet number ($\Rey\Pran$) plotted in figure~2($b$) for steady rolls of aspect ratio $\Gamma=2$ ($k=\pi$). The resolution of Fourier modes ($N_x$) and Chebyshev collocation points ($N_z$) is given also.}
\end{center}
\end{table}
\vspace*{\fill}

\newpage

\begin{table}[t!]
 \begin{center}
\begin{tabular}{ccccccc}
\hline
\multirow{2}{*}{$Ra$} & 	&	\multicolumn{2}{c}{Steady roll}	&		&  \multicolumn{2}{c}{DNS}  \\ 
   &	&	$Nu$	 	&	$Re$	&	 &  $Nu$  &  $Re$  \\ \hline
$10^7$  &		&	53.2089	&	515.276	&	&	48.53	&	488.77\\
$3\times10^7$  &		&	77.5286	&	1080.31	&	&	69.43	&	1016.46\\
$10^8$  &		&	116.711	&	2425.10	&	&	96.81	&	2202.08\\
$3\times10^8$  &		&	169.146	&	5063.63	&	&	134.00	&	4552.41\\
$10^9$  &		&	253.606	&	11332.6	&	&	198.01	&	10135.1\\
\hline
  \end{tabular}
  \caption*{Table S4: Comparison of $Nu$ and $Re$ between steady rolls with fixed aspect ratio $\Gamma=3.2$ and unsteady DNS by \citet{Wang2020} with the same mean aspect ratio. In both cases $\Pran=10$.}{\label{tab:SteadyvsDNS}}
 \end{center}
\end{table}
\vspace*{\fill}
\newpage



\floatsetup[figure]{style=plain,subcapbesideposition=top}  
\begin{figure}[h!]
\vspace{0.20in}
\begin{minipage}[t]{0.025\textwidth}
\vspace{0pt}
(a)
\end{minipage}
\begin{minipage}[t]{0.46\textwidth}
\vspace{0pt}
\includegraphics[height=1.85in]{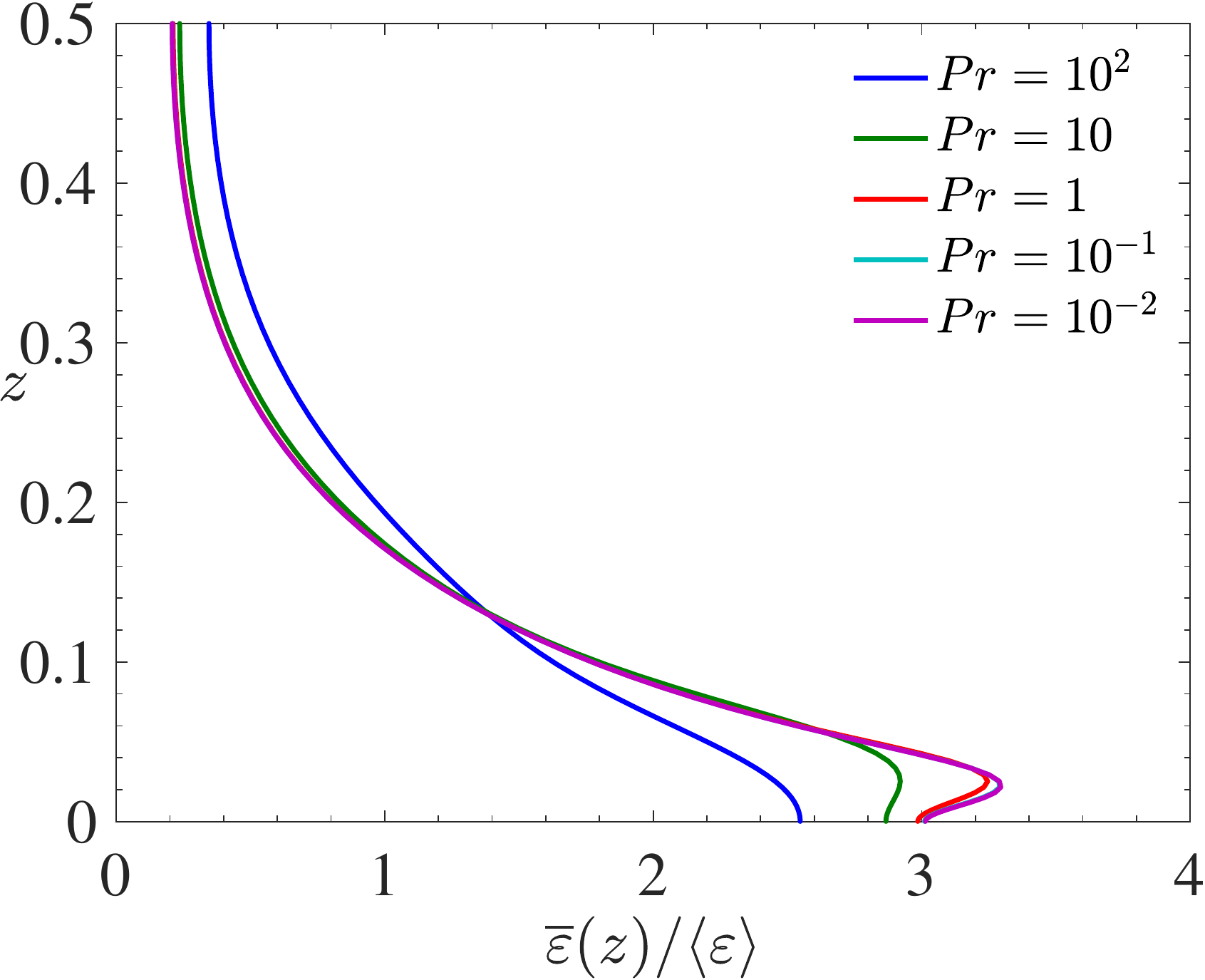}
\end{minipage}\hfill
\begin{minipage}[t]{0.035\textwidth}
\vspace{0pt}
(b)
\end{minipage}
\begin{minipage}[t]{0.46\textwidth}
\vspace{0pt}
\includegraphics[height=1.85in]{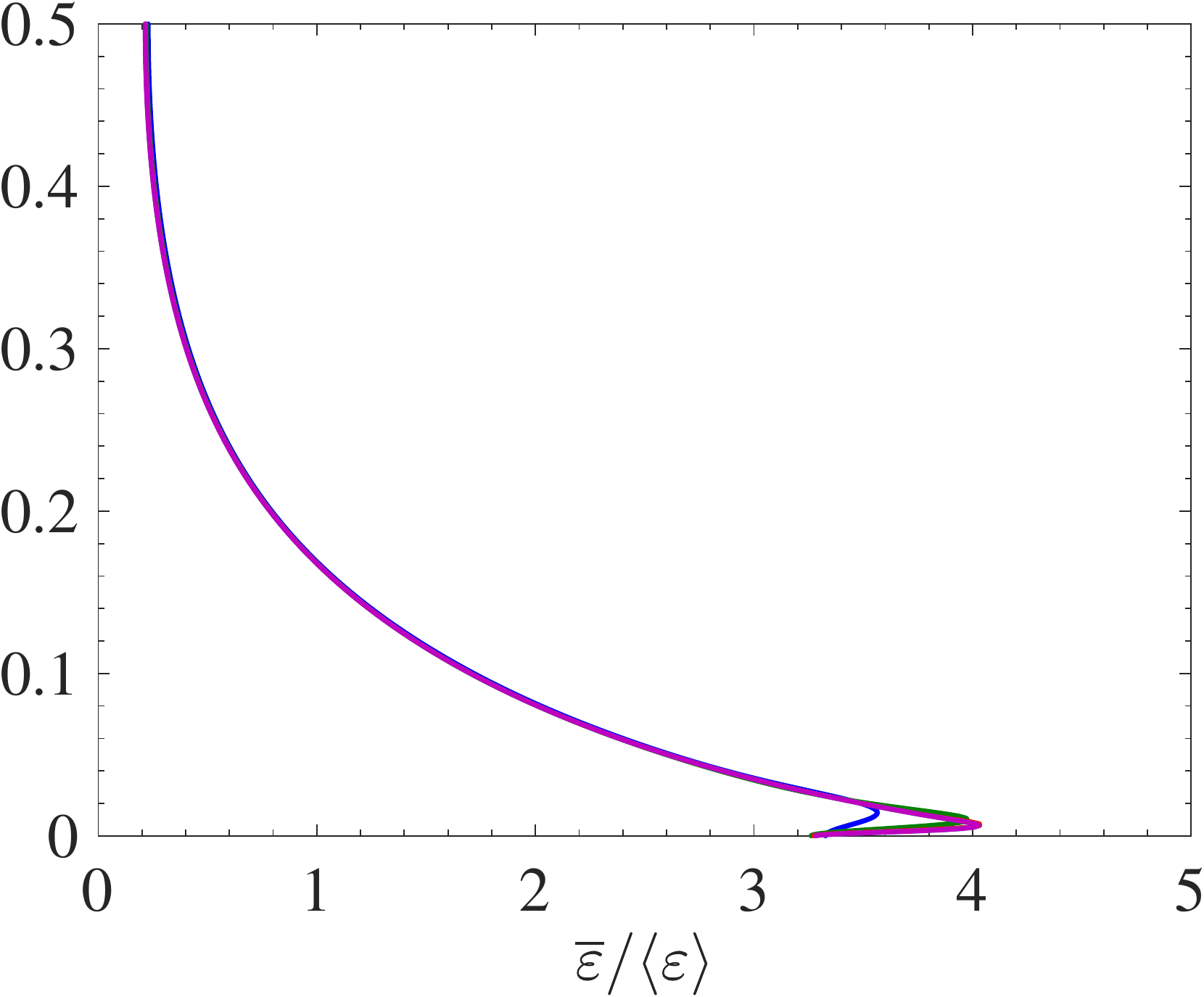}
\end{minipage}
\caption*{\label{fig: dissipation}Figure S1: Horizontally averaged dissipation profiles normalized by their volume averages, $\overline{\varepsilon}(z)/\langle \varepsilon \rangle$, for steady convective rolls at (a) $\Ra=5\times10^6$ and (b) $\Ra=10^{9}$ with $\Gamma=2$ and various $\Pran$. Only half of the vertical domain is shown ($0 \le z \le 0.5$) because the profiles are symmetric about the mid-plane.}
\end{figure}

\newpage

\bibliographystyle{jfm}
\bibliography{JFMRapids}

\end{document}